# Dynamical properties of the two-dimensional Holstein-Hubbard model in the normal state at zero temperature: A fluctuation-based effective cumulant approach

T. Hakioğlu
*Physics Department, Bilkent University, 06533 Ankara, Turkey*

M. Ye. Zhuravlev
*N. S. Kurnakov Institute of General and Inorganic Chemistry of the Russian Academy of Sciences,
31 Leninsky Prospect, GSP-1, 117 907 Moscow, Russia*

The two-dimensional many-body Holstein-Hubbard model in the $T=0$ normal state is examined within the framework of the self-consistent coupling of charge fluctuation correlations to the vibrational ones. The parameters of our model are the adiabaticity, the electron concentration, as well as the electron-phonon and the Coulomb interaction strengths. A fluctuation-based effective cumulant approach is introduced to examine the $T=0$ normal-state fluctuations and an analytic approximation to the true dynamical entangled ground state is suggested. Our results for the effective charge-transfer amplitude, the ground state energy, the fluctuations in the phonon population, the phonon softening as well as the coupling constant renormalizations suggest that, the recent numerical calculations of de Mello and Ranninger (Ref. 5), Berger, Valášek, and von der Linden (Ref. 2), and Marsiglio (Refs. 4 and 8) on systems with finite degrees of freedom can be qualitatively extended to the systems with large degrees of freedom. [S0163-1829(98)03728-X]

## I. INTRODUCTION

In this work we focus on the dynamical properties of the polaronic ground state in the Holstein-Hubbard model from the perspective of what we call as the charge-density wave (CDW) fluctuation-based *effective cumulant approach*. In this many-body model, the qualitative aspects of the transition from large to small polarons as the electron-phonon ($e$-ph) adiabaticity and the Coulomb interaction strengths are varied, with the full assessment of these interactions, is still an unresolved problem since the celebrated work of Holstein.[1] Recently quantum Monte Carlo (QMC) calculations,[2–4] semianalytic direct diagonalization[5–8] using finite lattice and electronic degrees of freedom, and variational ground-state techniques[9–11] have revealed evidence of a smooth transition of the ground state from the large extended to the small localized polaronic one as the interaction parameters are varied from the weak-coupling adiabatic to strong-coupling antiadiabatic ranges. The ground-state dynamics of the Holstein-Hubbard model is determined by the three dimensionless scales; viz., the adiabaticity $\gamma = t/\omega_0$, the $e$-ph mediated coupling $\lambda = (g/\omega_0)^2$, and the repulsive Coulomb interaction strength $V_c^{e-e}/\omega_0$ where $\omega_0$ is the frequency of Einstein phonons, $t$ is the charge transfer amplitude and $g$ is the linear $e$-ph coupling strength. In the weak-coupling adiabatic regime (i.e., $\lambda < 0.5$, $1 < \gamma$ and $V_c^{e-e}$ sufficiently small), the Migdal random-phase approximation (RPA) is quite accurate in describing the quasiparticle renormalization. However, the extension of Migdal RPA beyond $\lambda \sim 0.5$ encounters superficial instabilities in the phonon vacuum. This point has been critically questioned for instance, in Refs. 2, 5, and 6 when it is no longer possible to assign independent degrees of freedom to phonon and electron systems beyond the weak-coupling strongly adiabatic ranges, and one has to self-consistently deal with an entangled dynamical picture by abandoning the simpler quasiparticle one. On the semianalytic side progress has been made in the diagrammatic approaches by extending the Migdal RPA to the Migdal-Eliashberg (ME) formalism with self-consistent handling of the phonon and electron renormalizations within the RPA, where compatible results to more reliable QMC simulations[2,4] have been obtained. At the other extreme, the crucial role played by the adiabaticity parameter was clearly shown in Ref. 5 such that the strong-coupling Lang-Firsov (LF) approximation is strictly applicable only in the strongly antiadiabatic range $\gamma \sim 1$ and contrary to the common belief, the convergence to LF behavior can be considerably weakened in transition from strongly antiadiabatic $\gamma \ll 1$ to weakly nonadiabatic ranges $\gamma \lesssim 1$. In our opinion, although these results do not contradict the conditions of applicability of the LF approach or strong-coupling $1/\lambda$ expansion,[12] they confine their validity to the strongly antiadiabatic ranges.

The crucial point needed for a global perspective of the ground-state properties in the Holstein-Hubbard problem for a large range of coupling constants and adiabaticities is in the understanding of the nonlinear, self-consistent coupling of the charge fluctuations to the fluctuations in the vibrational degrees of freedom. In this respect, the main motivation for our fluctuation-based approach was provided by the numerical direct-diagonalization results in Ref. 5 regarding the correlated charge-deformation dynamics, as well as the CDW susceptibility based QMC and self-consistent ME calculations of Refs. 2, 4, and 7.

It is desirable that these numerical calculations, despite the limitations in the consideration of finite degrees of freedom—such as the finite lattice size, truncated Hilbert space, small number of electrons etc., which are necessary from the feasibility point of view of the numerical

methods—can be qualitatively extended to reach at conclusive results on the nature of the polaronic transition for more generalized systems with large degrees of freedom. In fact, this point has been raised a long time ago by Shore and Sander[9] and also stressed by the authors of Refs. 5, 7, and 8. However there has not been conclusive evidence, in particular at the intermediate ranges $\gamma \simeq 1$, $\lambda \simeq 1$, on whether the theoretical results obtained using models with finite degree of freedom could be extended to systems with realistic sizes. Moreover, apart from the variational ground state calculations,[9–11] direct attempts to tackle the many body dynamical fluctuations, in particular in the intermediate ranges, have not been possible on practical grounds. On the other hand, although the self-consistent ME RPA as well as QMC calculations provide an improved understanding of the problem, a clear self-consistent picture of fluctuations in the ground state (and, perhaps an approximate analytic form) still remains to be established.

In this work we approach the many-body problem in the normal state and at zero temperature by improving the CDW fluctuation-based effective cumulant approach that was recently introduced in Ref. 13 and applied to the superconducting-state solution to examine the low-temperature $T_c$-dependent phonon anomalies in certain high-temperature superconductors.

In Sec. II the Holstein-Hubbard model is introduced and studied in the momentum space. The nature of the interacting ground state is examined in Sec. II A where an approximate analytic form is suggested in the direct product form, decomposing the entangled nonlinear polaronic wave function in the coherent and two-particle correlated subspaces. The parameters of this effective wave function are calculated using the CDW fluctuation-based effective cumulant approach, reproducing the all first and second-order phonon cumulants of the entangled polaronic wave function. The effective wave function is an analytic and continuous function of $\lambda$, $\gamma$, and $V_c^{e-e}$, which ensures the same properties for the ground-state energy as well as other physical parameters induced from the model. The solution of the wave-function parameters as well as the calculation of the approximate ground-state energy is presented in Sec. II B. Section III is devoted to the renormalization of the charge-transfer amplitude. In Sec. IV, the renormalization of the effective $e$-$e$ interaction is examined. The statistics of the fluctuations in the ground state of the renormalized phonon subsystem and the renormalization of the vibrational frequency are examined in Secs. V A and V B, respectively.

## II. MODEL

We investigate the Holstein-Hubbard problem via the Hamiltonian,

$$\mathcal{H} = \mathcal{H}_e + \mathcal{H}_{\text{ph}} + \sum_{\mathbf{k},\mathbf{m},\sigma} g(\mathbf{k}) e^{i\mathbf{k}\mathbf{m}} c^\dagger_{\mathbf{m}\sigma} c_{\mathbf{m}\sigma} (a_\mathbf{k} + a^\dagger_{-\mathbf{k}})$$

$$+ \frac{1}{2} \sum_{\mathbf{m},\mathbf{n},\sigma,\sigma'} V_{\mathbf{m},\mathbf{n}} c^\dagger_{\mathbf{m}\sigma} c^\dagger_{\mathbf{n}\sigma'} c_{\mathbf{n}\sigma'} c_{\mathbf{m}\sigma}, \quad (1)$$

where $c^\dagger_{\mathbf{m}\sigma}, c_{\mathbf{m}\sigma}$ create and annihilate electrons at site $\mathbf{m}$ with spin $\sigma$ on a two-dimensional (2D) lattice, $a^\dagger_\mathbf{k}, a_\mathbf{k}$ create and annihilate phonons at momentum $\mathbf{k}$ with $g(\mathbf{k})$, and $V_{\mathbf{m},\mathbf{n}}$ describing the linear $e$-ph and electron-electron Coulomb interactions, respectively. The first two terms in the Hamiltonian describe the electron charge transfer and the harmonic phonon contributions as

$$\mathcal{H}_e = \sum_{\langle \mathbf{mn} \rangle \sigma} t_{\mathbf{mn}} c^\dagger_{\mathbf{m}\sigma} c_{\mathbf{n}\sigma}, \quad \text{and}$$

$$\mathcal{H}_{\text{ph}} = \sum_\mathbf{k} \frac{\omega_\mathbf{k}}{2} (a^\dagger_\mathbf{k} a_\mathbf{k} + a_\mathbf{k} a^\dagger_\mathbf{k}), \quad (2)$$

where $t_{\mathbf{m},\mathbf{n}}$ is the translationally invariant charge-transfer amplitude between neighboring sites $\mathbf{m}$, $\mathbf{n}$ and $\omega_\mathbf{k}$ is the harmonic phonon frequency.

The central theme of this work is to calculate the fluctuations in the vibrational degrees of freedom in a self-consistent frame together with the charge-density fluctuations in the correlated electron subsystem. The charge-density fluctuations are defined by the expressions

$$c^\dagger_{\mathbf{m}\sigma} c_{\mathbf{m}\sigma} = \langle c^\dagger_{\mathbf{m}\sigma} c_{\mathbf{m}\sigma} \rangle + \Delta \{c^\dagger_{\mathbf{m}\sigma} c_{\mathbf{m}\sigma}\} \quad \text{or, equivalently,}$$

$$\rho_\mathbf{k} = 2\bar{n}_\mathbf{k} + \delta\rho_\mathbf{k}, \quad (3)$$

where $\rho_\mathbf{k} = \Sigma_{\mathbf{k}',\sigma} c^\dagger_{\mathbf{k}'+\mathbf{k},\sigma} c_{\mathbf{k}',\sigma}$ with $c^\dagger_{\mathbf{k},\sigma}, c_{\mathbf{k},\sigma}$ describing the electron operators in the momentum representation, and $2\bar{n}_\mathbf{k} = \langle \rho_\mathbf{k} \rangle$ describing the CDW order parameter. The factor of 2 in Eq. (3) arises from the spin degeneracy. Using Eq. (3), Eq. (1) is separated into $\mathcal{H} = \mathcal{H}_0 + \mathcal{H}_I$ such that

$$\mathcal{H}_0 = \mathcal{H}_e + \sum_\mathbf{k} \left\{ \frac{\omega_\mathbf{k}}{2} (a^\dagger_\mathbf{k} a_\mathbf{k} + a_\mathbf{k} a^\dagger_\mathbf{k}) + 2g(\mathbf{k}) \bar{n}_\mathbf{k} (a_\mathbf{k} + a^\dagger_{-\mathbf{k}}) \right\},$$

$$\mathcal{H}_I = \sum_\mathbf{k} g(\mathbf{k}) \delta\rho_\mathbf{k} (a_\mathbf{k} + a^\dagger_{-\mathbf{k}}) + \frac{1}{2} \sum_\mathbf{k} V_c(\mathbf{k}) \rho_\mathbf{k} \rho_{-\mathbf{k}}, \quad (4)$$

where $V_c(\mathbf{k}) = 1/N \Sigma_\mathbf{k} e^{i\mathbf{k}\cdot(\mathbf{m}-\mathbf{n})} V_{\mathbf{m},\mathbf{n}}$, and $\mathcal{H}_0$ corresponds to the exactly solvable part associated with the eigen-wavefunction,

$$|\psi_0\rangle = |\phi_c\rangle \otimes |\psi_e\rangle,$$

$$|\phi_c\rangle = \mathcal{U}_c |0_{\text{ph}}\rangle = \exp\left\{ 2 \sum_\mathbf{k} \frac{g(\mathbf{k})}{\omega_\mathbf{k}} \bar{n}_\mathbf{k} (a_\mathbf{k} - a^\dagger_{-\mathbf{k}}) \right\} |0_{\text{ph}}\rangle. \quad (5)$$

Here $|\phi_c\rangle$ describes the pure coherent part of the ground-state wave function in the phonon subsystem, and $|0_{\text{ph}}\rangle$ is the phonon vacuum state. At the exactly solvable level the product form of the wave function remains to be valid with $|\psi_e\rangle$ representing the wave function of the electron subsystem. The coherent part $|\phi_c\rangle$ describes the coupling of the phonons to the static charge-density wave described by the CDW order parameter $\bar{n}_\mathbf{k} = 1/2 \Sigma_{\mathbf{k}',\sigma} \langle c^\dagger_{\mathbf{k}+\mathbf{k}',\sigma} c_{\mathbf{k}',\sigma} \rangle$. To examine the dynamical contributions to the interacting ground-state wave function we eliminate this part from the Hamiltonian by the unitary Lang-Firsov transformation $\mathcal{U}_c$ in Eq. (5) as

$$\mathcal{H}' = \mathcal{U}_c^\dagger(\mathcal{H}_0 + \mathcal{H}_I)\mathcal{U}_c$$

$$= \mathcal{H}_e + \sum_{\mathbf{k}} \frac{\omega_k}{2}(a_\mathbf{k}^\dagger a_\mathbf{k} + a_\mathbf{k} a_\mathbf{k}^\dagger)$$

$$- \sum_{\mathbf{k}} \frac{|g(\mathbf{k})|^2}{\omega_\mathbf{k}} (\bar{n}_\mathbf{k} + \delta\rho_\mathbf{k})\bar{n}_{-\mathbf{k}} + \mathcal{H}_I, \qquad (6)$$

where the coherent part of the wave function in Eq. (5) is now shifted to $|\psi_0'\rangle = \mathcal{U}_c^\dagger|\psi_0\rangle = |0_{\text{ph}}\rangle \otimes |\psi_e\rangle$. Now the interaction term $\mathcal{H}_I$ in Eq. (4) is given purely in terms of the coupling of phonons to the fluctuations in the CDW that contribute to the dynamical part of the interacting ground-state wave function. This interaction term is also conventionally transformed away by another unitary transformation $\mathcal{U}_\delta$,

$$\mathcal{U}_\delta = \exp\left\{\sum_\mathbf{k} \frac{g(\mathbf{k})}{\omega_\mathbf{k}} \delta\rho_\mathbf{k}(a_\mathbf{k} - a_{-\mathbf{k}}^\dagger)\right\}, \qquad (7)$$

for which the transformed Hamiltonian reads

$$\mathcal{H}'' = \mathcal{U}_\delta^\dagger \mathcal{H}' \mathcal{U}_\delta$$

$$= \sum_{\langle \mathbf{mn}\rangle\sigma} t_{\mathbf{mn}}\sigma(\mathbf{m},\mathbf{n})c_{\mathbf{m}\sigma}^\dagger c_{\mathbf{n}\sigma}$$

$$+ \sum_\mathbf{k} \frac{\omega_\mathbf{k}}{2}(a_\mathbf{k}^\dagger a_\mathbf{k} + a_\mathbf{k} a_\mathbf{k}^\dagger) - \sum_\mathbf{k} \frac{|g(\mathbf{k})|^2}{\omega_\mathbf{k}} \delta\rho_\mathbf{k}\delta\rho_{-\mathbf{k}}$$

$$- \sum_\mathbf{k} \frac{|g(\mathbf{k})|^2}{\omega_\mathbf{k}}(\bar{n}_\mathbf{k} + \delta\rho_\mathbf{k})\bar{n}_{-\mathbf{k}} + \frac{1}{2}\sum_\mathbf{k} V_c(\mathbf{k})\rho_\mathbf{k}\rho_{-\mathbf{k}}. \qquad (8)$$

The expense paid by this transformation is the introduction of the multiphonon operator,

$$\sigma(\mathbf{m},\mathbf{n}) = \exp\left[\frac{1}{2}\sum_\mathbf{k} \frac{g(\mathbf{k})}{\omega_\mathbf{k}}(e^{i\mathbf{k}\cdot\mathbf{m}} - e^{i\mathbf{k}\cdot\mathbf{n}})(a_\mathbf{k} - a_{-\mathbf{k}}^\dagger)\right]. \qquad (9)$$

Combining the transformations in Eq. (5) and Eq. (7) we obtain a highly entangled dynamical wave function $|\psi_\delta\rangle = \mathcal{U}_\delta^\dagger|\psi_0'\rangle$. Although the rest of the Hamiltonian in Eq. (8) is decoupled in electron and phonon degrees of freedom, a major difficulty is introduced by the multiphonon-electron scattering in the first term in Eq. (8). In the conventional Lang-Firsov approach this term is replaced by its average in the coherent part $|\phi_c\rangle$ of the wave function by $\sigma(\mathbf{m},\mathbf{n}) \rightarrow \langle\phi_c|\sigma(\mathbf{m},\mathbf{n})|\phi_c\rangle$, which completely decouples the Hamiltonian. On the other hand, a refined treatment of the residual interactions induced by $\sigma(\mathbf{m},\mathbf{n}) - \langle\phi_c|\sigma(\mathbf{m},\mathbf{n})|\phi_c\rangle$ has to incorporate the highly nonlinear phonon correlator,[12] which, in our opinion, can obscure the physical picture of the dynamical properties of the wave function.

In fact, the difficulties in the solution of the many-body problem are, at least, twofold. At one end, there is the impracticality of a formal diagrammatical approach to the residual interactions.[12] At the other end, even if one can get away with neglecting the residual interactions by using a Lang-Firsov-like formalism, a full understanding of the dynamical wave function $|\psi_\delta\rangle$ is still not promised due to its highly entangled nature. In this work, we will approach the problem within a Lang-Firsov-*like* approach (namely, by replacing $\sigma_{\mathbf{m},\mathbf{n}}$ by $\langle\psi_\delta|\sigma_{\mathbf{m},\mathbf{n}}|\psi_\delta\rangle$ where $|\psi_\delta\rangle$ is, in contrast to the LF approach where the coherent part is used, the dynamical fluctuating part of the polaron wave function) by demonstrating that it is, in principle, possible to construct an effective wave function $|\psi_\delta^{\text{eff}}\rangle$ as an approximation to the dynamical part $|\psi_\delta\rangle$, which adopts a special product form at the cumulant-generating-operator level in the $n$-phonon cumulant correlation space. Then, in an approximation scheme, an analytic form $|\psi_\delta^{\text{eff}}\rangle$ will be constructed by reproducing all first- and second-order cumulants of the phonon operators in $|\psi_\delta\rangle$.

### A. Nature of the interacting ground state

Our purpose in this subsection is to understand the nature of the dynamical strongly entangled wave function $|\psi_\delta\rangle$. In the static CDW limit (i.e., $\bar{n}_\mathbf{k} \neq 0$), the fluctuations in the charge density are negligible. It is known that the static CDW limit corresponds to strongly antiadiabatic regimes when the $e$-ph coupling constant is in the extreme weak- or strong-coupling limits. This is the limit where $|\phi_c\rangle$ can accurately approximate the exact polaron ground state of $\mathcal{H}$ in Eq. (1). In the weak-coupling antiadiabatic limit $\lambda \ll 1$, $\gamma \ll 1$, a perturbative scheme based on charge fluctuations is adequate where the magnitude of fluctuations in the residual interactions is limited [i.e., $|\sigma(\mathbf{m},\mathbf{n}) - \langle\sigma(\mathbf{m},\mathbf{n})\rangle| \ll 1$]; since $\sigma(\mathbf{m},\mathbf{n})$ is a positive and bounded operator by unity from above and $\langle\sigma(\mathbf{m},\mathbf{n})\rangle \simeq 1$. In the strong-coupling antiadiabatic regime, the small polaronic bandwidth is strongly reduced where we also have negligible contribution of the residual interactions. There, $\sigma(\mathbf{m},\mathbf{n})$ is bounded from below by zero since $\langle\sigma(\mathbf{m},\mathbf{n})\rangle \ll 1$. It is clear that the corrections to $|\phi_c\rangle$ as well as the importance of the residual interactions arise from the nonnegligible presence of the dynamical fluctuations in the intermediate ranges between these limits.

We will examine $|\psi_\delta\rangle$ by calculating the characteristic cumulants of the phonon coordinates $Q_\mathbf{k} = 1/\sqrt{2}(a_\mathbf{k} + a_{-\mathbf{k}}^\dagger)$ and $P_\mathbf{k} = -i/\sqrt{2}(a_{-\mathbf{k}} - a_\mathbf{k}^\dagger)$. In order to study the dynamical fluctuations in the ground state we shift the phonon coordinates in the Hamiltonian (1) to the origin by $Q_\mathbf{k} \rightarrow Q_\mathbf{k} - \langle Q_\mathbf{k}\rangle$ and $P_\mathbf{k} \rightarrow P_\mathbf{k} - \langle P_\mathbf{k}\rangle$ where $\langle Q_\mathbf{k}\rangle$ and $\langle P_\mathbf{k}\rangle$ are determined in the coherently shifted component $|\phi_c\rangle$ as $\langle Q_\mathbf{k}\rangle = 2[g(\mathbf{k})/\omega_\mathbf{k}]\bar{n}_\mathbf{k}$ and $\langle P_\mathbf{k}\rangle = 0$. This is equivalent to a unitary transformation by $\mathcal{U}_c$ of the initial Hamiltonian yielding Eq. (6). Note that from here on all expressions involving factors of $Q_\mathbf{k}$ and $P_\mathbf{k}$ will be expressed in the *shifted coordinates*.

We start with calculating five distinct types of the phonon moments defined by

$$\mathcal{R}_{s_1} = \langle\psi_\delta|(Q_\mathbf{k})^{s_1}|\psi_\delta\rangle,$$

$$\mathcal{P}_{s_2} = \langle\psi_\delta|(P_\mathbf{k})^{s_2}|\psi_\delta\rangle,$$

$$\mathcal{K}_{s_3} = \langle\psi_\delta|(P_\mathbf{k}P_{-\mathbf{k}})^{s_3}|\psi_\delta\rangle,$$

$$\mathcal{F}_{s_4} = \langle\psi_\delta|(Q_\mathbf{k}Q_{-\mathbf{k}})^{s_4}|\psi_\delta\rangle,$$

$$\mathcal{G}_{s_5} = \langle\psi_\delta|(Q_\mathbf{k})^{s_5}(P_\mathbf{k})^{s_5}|\psi_\delta\rangle. \qquad (10)$$

After a tedious but straightforward calculation using $|\psi_\delta\rangle = \mathcal{U}_\delta^\dagger|0_{\text{ph}}\rangle \otimes |\psi_e\rangle$, these are explicitly given by

$$\mathcal{R}_{s_1} = 0,$$

$$\mathcal{P}_{s_2} = 0,$$

$$\mathcal{K}_{s_3} = \left(\frac{1}{2}\right)^{s_3} s_3!,$$

$$\mathcal{F}_{s_4} = \frac{s_4!}{2^{s_4}} \sum_{p=0}^{\infty} {}_2F_1(-s_4+p,0;1;-1)$$
$$\times \frac{(-s_4)_p(-2p)_p}{(p!)^2}\left(\frac{g(\mathbf{k})}{\omega_\mathbf{k}}\right)^{2p} (\langle \delta\rho_\mathbf{k}\delta\rho_{-\mathbf{k}}\rangle)^p,$$

$$\mathcal{G}_{s_5} = \left(\frac{i}{2}\right)^{s_5} s_5!, \qquad (11)$$

where $(n)_m = n(n+1)\cdots(n+m-1)$ and ${}_2F_1(a,b;c;z)$ is the Gauss hypergeometric function and we assumed Gaussian density fluctuation correlations. In principle, an effective wave function $|\psi_\delta^{\text{eff}}\rangle$ that is expected to be equivalent to $|\psi_\delta\rangle$ in the phonon sector should consistently reproduce the entire set of an infinite number of cumulants in Eqs. (11) with $1 \leq s_i < \infty$, $(i=1,\ldots,5)$. Hence, the effective wave function also comprises an infinitely large set of correlation subspaces where the correlations in each subspace is produced by the unitary $n$-phonon cumulant correlation generator $\mathcal{U}^{(n)}$ as

$$|\psi_\delta^{\text{eff}}\rangle = \prod_{n=1}^{\infty} \mathcal{U}^{(n)}|0_{\text{ph}}\rangle \otimes |\psi_e\rangle,$$

where

$$\prod_{n=1}^{\infty} \mathcal{U}^{(n)} \equiv \prod_{n=m}^{\infty} \mathcal{U}^{(n)}\mathcal{U}^{(m-1)}\cdots\mathcal{U}^{(2)}\mathcal{U}^{(1)} \qquad (12)$$

with $\mathcal{U}^{(1)}, \mathcal{U}^{(2)}, \ldots$, etc. describing the one-particle coherent, the two-particle coherent correlations, etc., respectively. In fact, in this decomposition in terms of correlation subspaces, $\mathcal{U}^{(1)}$ corresponds to the coherent shift $\mathcal{U}_c^\dagger$ in Eq. (5) and $\mathcal{U}^{(2)}, \mathcal{U}^{(3)}$, etc. describe the two-particle and three-particle correlated sectors of $\mathcal{U}_\delta^\dagger$ in Eq. (7), etc. In this case the projection of the effective wave function $|\psi_\delta^{\text{eff}}\rangle$ on the $m$-dimensional correlation subspace is $|\psi_\delta\rangle_m$, which is determined by the projection operator,

$$\mathcal{T}_m = \left(\prod_{n=m+1}^{\infty} \mathcal{U}^{(n)}\right)^\dagger \quad \text{as} \quad |\psi_\delta\rangle_m = \mathcal{T}_m|\psi_\delta\rangle. \qquad (13)$$

In order for the product form in Eq. (12) to be a sensible expansion of the wave function in terms of its independent sectors in the correlation space, each unitary $n$-phonon correlation generator $\hat{\mathcal{U}}^{(n)}$ must reproduce the $n$th-order phonon cumulants obtained from the moments in Eqs. (10) but not the moments themselves. This is indeed the reason why we shifted the phonon coordinates in order to eliminate the influence of the coherent one-particle sector on the second-order and higher dynamic correlations in the wave function. This is equivalent to subtracting the coherent one-particle contributions by performing the shift $Q_\mathbf{k} \rightarrow Q_\mathbf{k} - \langle 0_{\text{ph}}|\hat{\mathcal{U}}_c^\dagger Q_\mathbf{k} \hat{\mathcal{U}}_c|0_{\text{ph}}\rangle$. For those of the $m$th-order ones, this procedure defines generalized shifts $\langle 0_{\text{ph}}|(\Pi_{n=1}^{m-1}\hat{\mathcal{U}}^{(n)})^\dagger Q_\mathbf{k}(\Pi_{n'=1}^{m-1}\hat{\mathcal{U}}^{(n')})|0_{\text{ph}}\rangle$ and similarly for $P_\mathbf{k}$. In result, it is technically possible to decompose the wave function in direct product form in the cumulant correlation space. Despite the fact that the technical principles of such a decomposition prescribed in Eq. (12) can be examined, it is not practically possible to go beyond the second-order correlations, because of the fact that a possible general analytic form for the third- and higher-order cumulant generators $\hat{\mathcal{U}}^{(m)}$, $(3 \leq m)$ have not been studied in the literature from the mathematical point of view. The first- and the second-order cumulant correlations, on the other hand, are well known in quantum optics as the one-particle coherent[14] and the two-particle coherent states,[15-17] respectively, and have been extensively applied to the polaron problem in the context of the dynamical[13] and the variational (see, for instance, Refs. 18 and 19) approaches.

Under these practical limitations arising for $3 \leq m$, we consider a subset of Eqs. (10) comprising the entire first- and second-order cumulants, which correspond to $s_1, s_2 = 1, 2$, $s_3, s_4, s_5 = 1$. Hence, it is implied that the polaron ground-state wave function will be approximated in the cumulant correlation space using only the first- and the second-order cumulants. From Eq. (11) these seven cumulants are explicitly given by

$$\langle \psi_\delta|Q_\mathbf{k}|\psi_\delta\rangle = \langle \psi_\delta|P_\mathbf{k}|\psi_\delta\rangle = \langle \psi_\delta|Q_\mathbf{k}Q_\mathbf{k}|\psi_\delta\rangle$$
$$= \langle \psi_\delta|P_\mathbf{k}P_\mathbf{k}|\psi_\delta\rangle = 0,$$

$$\langle \psi_\delta|Q_\mathbf{k}Q_{-\mathbf{k}}|\psi_\delta\rangle = 1/2\left[1 + 4\left|\left(\frac{g(\mathbf{k})}{\omega_\mathbf{k}}\right)\right|^2 \langle \delta\rho_\mathbf{k}\delta\rho_{-\mathbf{k}}\rangle\right],$$

$$\langle \psi_\delta|P_\mathbf{k}P_{-\mathbf{k}}|\psi_\delta\rangle = 1/2,$$

$$\langle \psi_\delta|Q_\mathbf{k}P_\mathbf{k}|\psi_\delta\rangle = i/2. \qquad (14)$$

In order to reproduce Eqs. (14), we propose the effective wave function,

$$|\psi_\delta^{\text{eff}}\rangle = \mathcal{S}(\{\xi\})\prod_\mathbf{k} \{\alpha_\mathbf{k} + \gamma_\mathbf{k}a_\mathbf{k}^\dagger a_{-\mathbf{k}}^\dagger + \beta_\mathbf{k}(a_\mathbf{k}^\dagger)^2(a_{-\mathbf{k}}^\dagger)^2\}|0_{\text{ph}}\rangle$$
$$\otimes |\psi_e\rangle, \qquad (15)$$

where the phonon coordinates are coherently shifted for the calculation of second-order correlations according to the procedure outlined above. The wave function is normalized as $|\alpha_\mathbf{k}|^2 + |\beta_\mathbf{k}|^2 + |\gamma_\mathbf{k}|^2 = 1$, where we neglect the overall phase of $|\psi_\delta^{\text{eff}}\rangle$ by considering $\alpha_\mathbf{k} = \bar{\alpha}_\mathbf{k}$, and

$$\mathcal{S}(\{\xi\}) = \exp\left\{-\sum_\mathbf{k} (\xi_\mathbf{k}a_\mathbf{k}a_{-\mathbf{k}} - \bar{\xi}_\mathbf{k}a_\mathbf{k}^\dagger a_{-\mathbf{k}}^\dagger)\right\},$$

$$\xi_\mathbf{k} = |\xi_\mathbf{k}|e^{i2\theta_\mathbf{k}}, \quad \mathcal{S}^\dagger = \mathcal{S}^{-1} \qquad (16)$$

describes the two-particle coherent, translationally invariant unitary operator (squeezing operator in quantum optics[15,16]). The unitary transformation defined by $\mathcal{S}(\{\xi\})$ on the phonon coordinates is given by

$$\mathcal{S}^\dagger(\{\xi\})Q_\mathbf{k}\mathcal{S}(\{\xi\}) = [\kappa_\mathbf{k} + \mathrm{Re}\{\mu_\mathbf{k}\}]Q_\mathbf{k} + \mathrm{Im}\{\mu_\mathbf{k}\}P_{-\mathbf{k}},$$

$$\mathcal{S}^\dagger(\{\xi\})P_\mathbf{k}\mathcal{S}(\{\xi\}) = [\kappa_\mathbf{k} - \mathrm{Re}\{\mu_\mathbf{k}\}]P_\mathbf{k} + \mathrm{Im}\{\mu_\mathbf{k}\}P_{-\mathbf{k}}, \quad (17)$$

where $\kappa_\mathbf{k} = \cosh(2|\xi_\mathbf{k}|)$ and $\mu_\mathbf{k} = e^{-i2\theta_\mathbf{k}}\sinh(2|\xi_\mathbf{k}|)$ such that $|\kappa_\mathbf{k}|^2 - |\mu_\mathbf{k}|^2 = 1$ as imposed by the unitarity of $\mathcal{S}(\{\xi\})$.

Using the wave function in Eq. (15) and the properties in Eqs. (17) we obtain

$$\langle\psi_\delta^{\mathrm{eff}}|Q_\mathbf{k}|\psi_\delta^{\mathrm{eff}}\rangle = 0,$$

$$\langle\psi_\delta^{\mathrm{eff}}|P_\mathbf{k}|\psi_\delta^{\mathrm{eff}}\rangle = 0,$$

$$\langle\psi_\delta^{\mathrm{eff}}|Q_\mathbf{k}Q_\mathbf{k}|\psi_\delta^{\mathrm{eff}}\rangle = 0,$$

$$\langle\psi_\delta^{\mathrm{eff}}|P_\mathbf{k}P_\mathbf{k}|\psi_\delta^{\mathrm{eff}}\rangle = 0,$$

$$\langle\psi_\delta^{\mathrm{eff}}|Q_\mathbf{k}Q_{-\mathbf{k}}|\psi_\delta^{\mathrm{eff}}\rangle = \mathrm{Re}\{\alpha_\mathbf{k}\gamma_\mathbf{k}(\kappa_\mathbf{k}+\mu_\mathbf{k})^2 + 2\beta_\mathbf{k}\bar\gamma_\mathbf{k}(\kappa_\mathbf{k}+\mu_\mathbf{k})^2\} + \frac{1}{2}(\alpha_\mathbf{k}^2 + 3|\gamma_\mathbf{k}|^2 + 5|\beta_\mathbf{k}|^2)|\kappa_\mathbf{k}+\mu_\mathbf{k}|^2,$$

$$\langle\psi_\delta^{\mathrm{eff}}|P_\mathbf{k}P_{-\mathbf{k}}|\psi_\delta^{\mathrm{eff}}\rangle = -\mathrm{Re}\{\alpha_\mathbf{k}\gamma_\mathbf{k}(\kappa_\mathbf{k}-\mu_\mathbf{k})^2 + 2\beta_\mathbf{k}\bar\gamma_\mathbf{k}(\kappa_\mathbf{k}-\mu_\mathbf{k})^2\} + \frac{1}{2}(\alpha_\mathbf{k}^2 + 3|\gamma_\mathbf{k}|^2 + 5|\beta_\mathbf{k}|^2)|\kappa_\mathbf{k}-\mu_\mathbf{k}|^2,$$

$$\langle\psi_\delta^{\mathrm{eff}}|Q_\mathbf{k}P_\mathbf{k}|\psi_\delta^{\mathrm{eff}}\rangle = \frac{i}{2}\{1 + (\bar\kappa_\mathbf{k}\mu_\mathbf{k} - \kappa_\mathbf{k}\bar\mu_\mathbf{k})(\alpha_\mathbf{k}^2 + 3|\gamma_\mathbf{k}|^2 + 5|\beta_\mathbf{k}|^2)\} - \mathrm{Im}\{(\gamma_\mathbf{k}\alpha_\mathbf{k} + 2\beta_\mathbf{k}\bar\gamma_\mathbf{k})(\bar\kappa_\mathbf{k}^2 - \bar\mu_\mathbf{k}^2)\}. \quad (18)$$

The parameters $\alpha_\mathbf{k}, \gamma_\mathbf{k}, \beta_\mathbf{k}, \kappa_\mathbf{k}, \mu_\mathbf{k}$ are determined by demanding the equality of Eqs. (18) and Eqs. (14). In fact, independently from specific values of $\alpha_\mathbf{k}$, $\gamma_\mathbf{k}$, $\beta_\mathbf{k}$, and $\xi_\mathbf{k}$, the effective wave function $|\psi_\delta^{\mathrm{eff}}\rangle$ satisfies a larger set of cumulants than given by the subset in Eqs. (18). First of all, the first two conditions on $\mathcal{R}_{s_1}$ and $\mathcal{P}_{s_2}$ in Eq. (11) are very strict, corresponding to the translational invariance of $|\psi_\delta\rangle$. These are also respected for all $s_1, s_2$ by $|\psi_\delta^{\mathrm{eff}}\rangle$ independently from $\alpha_\mathbf{k}$, $\gamma_\mathbf{k}$, $\beta_\mathbf{k}$ and $\xi_\mathbf{k}$. Furthermore, we also have $\langle\psi_\delta|(Q_\mathbf{k})^{s_5}(P_\mathbf{k})^{s_5}|\psi_\delta\rangle = \langle\psi_\delta^{\mathrm{eff}}|(Q_\mathbf{k})^{s_5}(P_\mathbf{k})^{s_5}|\psi_\delta^{\mathrm{eff}}\rangle = (i/2)^{s_5}s_5!$ for all $s_5$ and for all *arbitrary but real* $\alpha_\mathbf{k}$, $\gamma_\mathbf{k}$, $\beta_\mathbf{k}$, and $\xi_\mathbf{k}$. Hence, we are motivated to find a solution where the parameters are all real. Here, we switch to the polar coordinates $\beta_\mathbf{k} = |\beta_\mathbf{k}|\exp(i\theta_\beta)$, and similarly for the other parameters. From the last equations in Eqs. (18) and (14), we infer that $\theta_\kappa - \theta_\mu = m\pi$ with $m = 0, 1$ and $\mathrm{Im}\{\gamma_\mathbf{k}\alpha_\mathbf{k} + 2\beta_\mathbf{k}\bar\gamma_\mathbf{k}\} = 0$. For real parameters this trivially implies $|\gamma_\mathbf{k}||\alpha_\mathbf{k}|\sin\theta_\gamma = -2|\beta_\mathbf{k}||\gamma_\mathbf{k}|\sin(\theta_\beta - \theta_\gamma) = 0$, hence, $\theta_\gamma = r\pi$ ($r = 0, 1$), and $\theta_\beta = n\pi$ ($n = 0, 1$). With these conditions, there are five real equalities in the simultaneous solution of Eqs. (18) and (14) and four conditions (including two normalization conditions) to be satisfied. We consider the fifth condition as the minimization of the ground-state energy. Since all parameters are now real, we drop the absolute value signs, i.e., $|\alpha_\mathbf{k}| \to \alpha_\mathbf{k}$ and similarly for the others. We now have an effective wave function that respects the strict conditions imposed by the translational invariance indicated by $\mathcal{R}_{s_1}$ and $\mathcal{P}_{s_2}$ as well as the last condition indicated by $\mathcal{G}_{s_5}$ in Eqs. (10) at all orders. Consistency between Eqs. (18) and (14) now implies

$$\kappa_\mathbf{k}^2 - \mu_\mathbf{k}^2 = 1,$$

$$\alpha_\mathbf{k}^2 + \beta_\mathbf{k}^2 + \gamma_\mathbf{k}^2 = 1,$$

$$\left\{(-1)^r\gamma_\mathbf{k}[\alpha_\mathbf{k} + 2(-1)^n\beta_\mathbf{k}] + \frac{1}{2}(\alpha_\mathbf{k}^2 + 3\gamma_\mathbf{k}^2 + 5\beta_\mathbf{k}^2)\right\}$$
$$\times [\kappa_\mathbf{k} + (-1)^m\mu_\mathbf{k}]^2$$
$$= 1/2\left[1 + 4\left|\left(\frac{g(\mathbf{k})}{\omega_\mathbf{k}}\right)\right|^2\langle\delta\rho_\mathbf{k}\delta\rho_{-\mathbf{k}}\rangle\right],$$

$$\left\{(-1)^{r+1}\gamma_\mathbf{k}[\alpha_\mathbf{k} + 2(-1)^n\beta_\mathbf{k}] + \frac{1}{2}(\alpha_\mathbf{k}^2 + 3\gamma_\mathbf{k}^2 + 5\beta_\mathbf{k}^2)\right\}$$
$$\times [\kappa_\mathbf{k} - (-1)^m\mu_\mathbf{k}]^2 = 1/2. \quad (19)$$

This set of four equations will be closed by one additional constraint from the ground-state minimization, which we address in the following section.

### B. Solution of the parameters and approximations to the true ground-state energy

We now define the ground-state energy of the Hamiltonian in Eq. (8) by

$$E_0 = \langle \psi_\delta^{\text{eff}} | \mathcal{H}'' | \psi_\delta^{\text{eff}} \rangle$$

$$= \sum_{\mathbf{m},\mathbf{n}} t_{\mathbf{m},\mathbf{n}} \langle \sigma_{\mathbf{m},\mathbf{n}} \rangle \langle c_\mathbf{m}^\dagger c_\mathbf{n} \rangle + \sum_{\mathbf{k}} V_0(\mathbf{k}) \langle \delta\rho_\mathbf{k} \delta\rho_{-\mathbf{k}} \rangle$$

$$+ \sum_{\mathbf{k}} V_0(\mathbf{k}) \bar{n}_\mathbf{k} \bar{n}_{-\mathbf{k}} + \sum_{\mathbf{k}} \frac{\omega_\mathbf{k}}{2} [(|\kappa_\mathbf{k}|^2 + |\mu_\mathbf{k}|^2)$$

$$\times (1 + 3|\gamma_\mathbf{k}|^2 + 8|\beta_\mathbf{k}|^2) + 4\alpha_\mathbf{k}|\gamma_\mathbf{k}||\kappa_\mathbf{k}||\mu_\mathbf{k}|$$

$$\times \text{Re}\{e^{i(\theta_\kappa + \theta_\mu - \theta_\gamma)}\} + 8|\gamma_\mathbf{k}||\beta_\mathbf{k}||\kappa_\mathbf{k}||\mu_\mathbf{k}|$$

$$\times \text{Re}\{e^{i(\theta_\kappa + \theta_\mu + \theta_\gamma - \theta_\beta)}\}], \quad (20)$$

where $V_0(\mathbf{k}) = \frac{1}{2} V_c(\mathbf{k}) - \{|g(\mathbf{k})|^2 / \omega_\mathbf{k}\}$, is the bare $e$-$e$ interaction, and, the last sum in Eq. (20) is the result of $\Sigma_\mathbf{k} \omega_\mathbf{k}/2 \langle \psi_\delta^{\text{eff}} | (a_\mathbf{k}^\dagger a_\mathbf{k} + a_\mathbf{k} a_\mathbf{k}^\dagger) | \psi_\delta^{\text{eff}} \rangle$. The contribution from the multiphonon operator is more tedious to calculate, for which we obtain

$$\langle \psi_\delta^{\text{eff}} | \sigma_{\mathbf{m},\mathbf{n}} | \psi_\delta^{\text{eff}} \rangle = \prod_\mathbf{k} \exp(-A_\mathbf{k}) [\alpha_\mathbf{k}^2 + |\gamma_\mathbf{k}|^2 (1 + A_\mathbf{k})^2$$

$$+ |\beta_\mathbf{k}|^2/4 (4 - 12 A_\mathbf{k}^2 + A_\mathbf{k}^4) + 2 A_\mathbf{k} \text{Re}\{\alpha_\mathbf{k} \bar{\gamma}_\mathbf{k}\}$$

$$+ A_\mathbf{k}^2 \text{Re}\{\alpha_\mathbf{k} \bar{\beta}_\mathbf{k}\} + A_\mathbf{k}(2 - A_\mathbf{k})^2 \text{Re}\{\gamma_\mathbf{k} \bar{\beta}_\mathbf{k}\}], \quad (21)$$

with $A_\mathbf{k} = \frac{1}{2} [g(\mathbf{k}/\omega_\mathbf{k})]^2 e^{-4\xi_\mathbf{k}} (1 - \cos \mathbf{k}_x \cdot a - \cos \mathbf{k}_y \cdot a)$ where $a$ describes the lattice constant, which we take to be unity. For the lowest possible energy we must satisfy in Eq. (20)

$$\pi = \theta_\kappa + \theta_\mu - \theta_\gamma,$$

$$\pi = \theta_\kappa + \theta_\mu + \theta_\gamma - \theta_\beta,$$

$$|\beta_\mathbf{k}| = \frac{1}{2} |\gamma_\mathbf{k}||\kappa_\mathbf{k}||\mu_\mathbf{k}| \bigg/ (|\kappa_\mathbf{k}|^2 + |\mu_\mathbf{k}|^2), \quad (22)$$

where the last one in Eqs. (22) is obtained by minimizing the phonon part in Eq. (20) with respect to $|\beta_\mathbf{k}|$. The first two yield $\theta_\beta - 2\theta_\gamma = 0$, thus $\theta_\beta = 0$. Using this as well as $\theta_\gamma = r\pi$ obtained previously we find two possible solutions

$$\theta_\gamma = 0, \quad \theta_\beta = 0, \quad \theta_\kappa = 0, \quad \theta_\mu = \pi, \quad \text{and}$$

$$\theta_\gamma = \pi, \quad \theta_\beta = 0, \quad \theta_\kappa = 0, \quad \theta_\mu = 0. \quad (23)$$

Since the phases are all fixed, we turn to the calculation of the density fluctuation correlations. The ground-state energy in Eq. (20), as well as the parameters of the wave function in Eqs. (19) and (22) are functions of $\langle \delta\rho_\mathbf{k} \delta\rho_{-\mathbf{k}} \rangle$, which we determine using the dielectric function $\epsilon(\mathbf{k}, \omega)$ formalism as

$$V_0(\mathbf{k}) \langle \delta\rho_\mathbf{k} \delta\rho_{-\mathbf{k}} \rangle = - \int_0^\infty \frac{d\omega}{\pi} \text{Im}\left(\frac{1}{\epsilon(\mathbf{k}, \omega)}\right). \quad (24)$$

In the RPA, $\epsilon(\mathbf{k}, \omega)$ is given by

$$\epsilon(\mathbf{k}, \omega) = 1 - \frac{V_0(\mathbf{k}) P(\mathbf{k}, \omega)}{1 + V_0(\mathbf{k}) P(\mathbf{k}, \omega)}. \quad (25)$$

The electron polarization $P(\mathbf{k}, \omega)$ is obtained in the standard formulation by

$$P(\mathbf{k}, \omega) = 2 \sum_\mathbf{p} \frac{\theta[-\xi_\mathbf{p} - \Sigma_\mathbf{p}] - \theta[\xi_{\mathbf{p}+\mathbf{k}} + \Sigma_{\mathbf{p}+\mathbf{k}}]}{\omega + \xi_\mathbf{p} - \xi_{\mathbf{p}+\mathbf{k}} + \Sigma_\mathbf{p} - \Sigma_{\mathbf{p}+\mathbf{k}} + i\delta}, \quad (26)$$

where $\xi_\mathbf{k} = t_{\text{eff}}(\mathbf{k}) - \mu$, $t_{\text{eff}} = 2t\langle\sigma\rangle(1 - \cos k_x - \cos k_y)$. Since $\alpha_\mathbf{k}$, $\beta_\mathbf{k}$, $\gamma_\mathbf{k}$, and $\xi_\mathbf{k}$ are not determined at this level, we consider in $t_{\text{eff}}$, the zeroth-order approximation where we replace $\langle\sigma\rangle$ by its LF limit $\langle\sigma\rangle_{\text{LF}} = \exp\{-1/2 |g(\mathbf{k})|^2/\omega_k^2\}$. The chemical potential $\mu$ is fixed self-consistently by the zero-temperature constraint,

$$n_0 = \sum_\mathbf{k} \theta[-\xi_\mathbf{k} - \Sigma(\mathbf{k})], \quad (27)$$

with $\Sigma(\mathbf{k}) = -\Sigma_{\mathbf{k}'} \mathcal{V}(\mathbf{k}' - \mathbf{k}) \theta[-\xi_{\mathbf{k}'} - \Sigma(\mathbf{k}')]$ describing the exchange contribution to one particle energy renormalization. Since we are confined here to zero-temperature formalism, $\Sigma(\mathbf{k})$ is independent from $\mathbf{k}$ and just renormalizes the chemical potential. Hence the exchange contribution is ineffective in the denominator of Eq. (26).

### 1. Density fluctuation correlations

We obtain the solution Eqs. (24–27) numerically in two dimensions using Einstein phonons $\omega_\mathbf{k} = \omega_0$ and $\mathbf{k}$ independent dimensionless bare $e$-ph coupling $\lambda = (g/\omega_0)^2$. All energies are normalized by $\omega_0$. The dependence of $\langle \delta\rho_\mathbf{k} \delta\rho_{-\mathbf{k}} \rangle$ on the dimensionless parameters $\lambda$, $\gamma$, and $V_c(\pi,\pi)/\omega_0$ is shown in Figs. 1(a–c) at $\mathbf{k} = (\pi, \pi)$ and at half-filling, for the values $V_c(\pi,\pi)/\omega_0 = 0, 1, 2, 3, 4$, and $\gamma = 0.05, 0.1, 1$ with $0 \leq \lambda \leq 2$. In each curve the solid line, open circles, open triangles, solid circles, and solid triangles represent values of $V_c(\pi,\pi)/\omega_0$ as, respectively, indicated above. A quantitative comparison of the figures for a fixed Coulomb interaction strength indicates that, as the adiabaticity $\gamma$ is decreased, there is an overall suppression in the magnitude of the fluctuation correlations. This effect is also enhanced further by strong $e$-ph coupling particularly in the strongly antiadiabatic (i.e., $\gamma \ll 1$) ranges. On the other hand, as $\gamma$ increases towards the adiabatic range, correlations gradually increase for stronger $e$-ph coupling. This picture qualitatively agrees with the results obtained by direct-diagonalization calculations on finite systems where a cooperation is observed in the antiadiabatic range between the decreasing adiabaticity and the increasing coupling constant to suppress the quantum fluctuations. The overall effect of the increasing repulsive Coulomb interaction is to overcome the phonon-induced polaron attraction, which amounts to suppressing the fluctuations for small couplings and enhancing them in the strong-coupling ranges. At this level, we solve Eqs. (19) and (22) for the parameters of the effective wave function before we calculate the ground-state energy.

### 2. Parameters of the effective wave function

Once fluctuation correlations are determined, the phonon effective ground-state parameters $\alpha_\mathbf{k}$, $\gamma_\mathbf{k}$, $\beta_\mathbf{k}$, and $\xi_\mathbf{k}$ can be calculated from Eqs. (19, 22) for two branches as characterized by Eqs. (23). The solutions corresponding to these two branches are identical for $\alpha_\mathbf{k}$, $\beta_\mathbf{k}$, and $\gamma_\mathbf{k}$ and only differ very slightly for $\kappa_\mathbf{k}$ and $\mu_\mathbf{k}$. In this subsection, we only present the results for the first branch, whereas, both solutions will be explicitly used in the calculation of the approximate ground-state energy. In Figs. 2(a–d) the parameters of the effective wave function are plotted for $\mathbf{k}=(\pi,\pi)$ at half-filling in the same $\lambda$ range as in Fig. 1. As the $e$-ph interaction is increased, a strong competition is observed between the strengths of the pure two-particle coherent component given by $\alpha_{\pi,\pi}$ and the pair excitations on this state given by $\gamma_{\pi,\pi}$. In the intermediate ranges of the $e$-ph coupling (i.e., $\lambda \lesssim 1$), the pair excitation strength becomes comparable to the strength of the underlying two particle coherent component. The four particle excitation given by $\beta_{\pi,\pi}$ is limited in strength in the whole $\lambda$ range. On the other hand, Fig. 2(d) represents the parameters within the two-particle coherent component. For increasing $e$-ph interaction a rapid reduction is observed in $\exp(-2|\xi_{\pi,\pi}|)$. We observe that, because of the non-negligible strength of $\gamma_{\pi,\pi}$, the whole picture here is quite contrary to the common practice of replacing the effective phonon ground state by a variational pure two-particle coherent (squeezed) component (in which case we would have $\alpha_\mathbf{k} \equiv 1$, $\gamma_\mathbf{k}=\beta_\mathbf{k}\equiv 0$ for all $\mathbf{k}$) in the intermediate and strong coupling regimes. In Figs. 3(a–d) the same parameters are calculated for $\gamma=0.05$. As the system is shifted to increasingly antiadiabatic ranges (i.e., $\gamma \ll 1$), the relative strength $\alpha_{\pi,\pi}$ of the pure two-particle coherent component is approximately maintained in the entire coupling range with respect to the two- and four-particle correlated excitations represented by $\gamma_{\pi,\pi}$ and $\beta_{\pi,\pi}$, respectively. Hence, in this range of the interaction parameters, the two-particle coherent component $\alpha_{\pi,\pi}$ dominates the wave function where the two- and four-particle correlated excitations $\gamma_{\pi,\pi}$ and $\beta_{\pi,\pi}$ compete only with each other. Within the two-particle coherent component [as indicated in Fig. 3(d)] there is a also an increasing tendency to overlap with the conventional phonon vacuum. Nevertheless, we observe that $\exp(-2|\xi_{\pi,\pi}|)$ saturates around 70%, implying that the overlap with the vacuum does not exceed 30% [see Fig. 3(d)] even for such a strong antiadiabaticity as $\gamma=0.05$. Note that, a strong overlap of the dynamical part $|\psi_\delta^{\text{eff}}\rangle$ with the vacuum would indicate that the coherent part $|\phi_c\rangle$ is dominating the ground-state wave function. These results are in qualitative agreement with the direct diagonalization results of Ref. 5 where the observed convergence of the true ground state to the Lang-Firsov small polaron limit (indicated by the pure coherent part $|\phi_c\rangle$) is weaker than expected and strongly adiabaticity dependent. As the system is driven to even more antiadiabatic ranges, the charge fluctuations reduce their overall amplitude as the fluctuating component $|\psi_\delta^{\text{eff}}\rangle$ of the polaron wave function develops an ever increasing overlap with the conventional vacuum [i.e., as implied by the saturation in $\alpha_{\pi,\pi}$ at approximately 90% with $\gamma_{\pi,\pi},\beta_{\pi,\pi}$ saturating at limited strengths as well as the tendency of $\exp(-2|\xi_{\pi,\pi}|)$ to stay closer to unity in Fig. 3(d)]. Hence the ground-state polaron wave function gradually becomes more coherent and localized; nevertheless, we also observe that the convergence to this limit is weaker than conventionally expected.

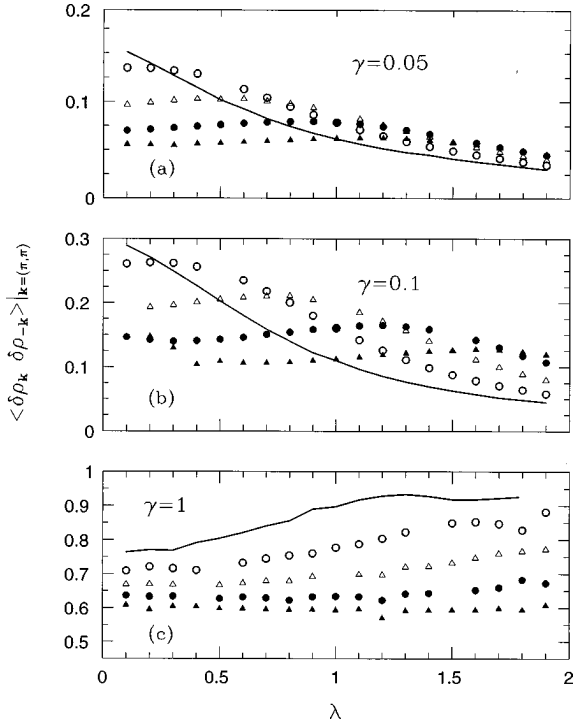

FIG. 1. The solution of the density fluctuation correlations at $\mathbf{k}=(\pi,\pi)$ and at half-filling as a function of the $e$-ph coupling for $V_c/\omega_0=0,1,2,3,4$ as represented by solid line, open circles, open triangles, solid circles, and solid triangles, respectively.

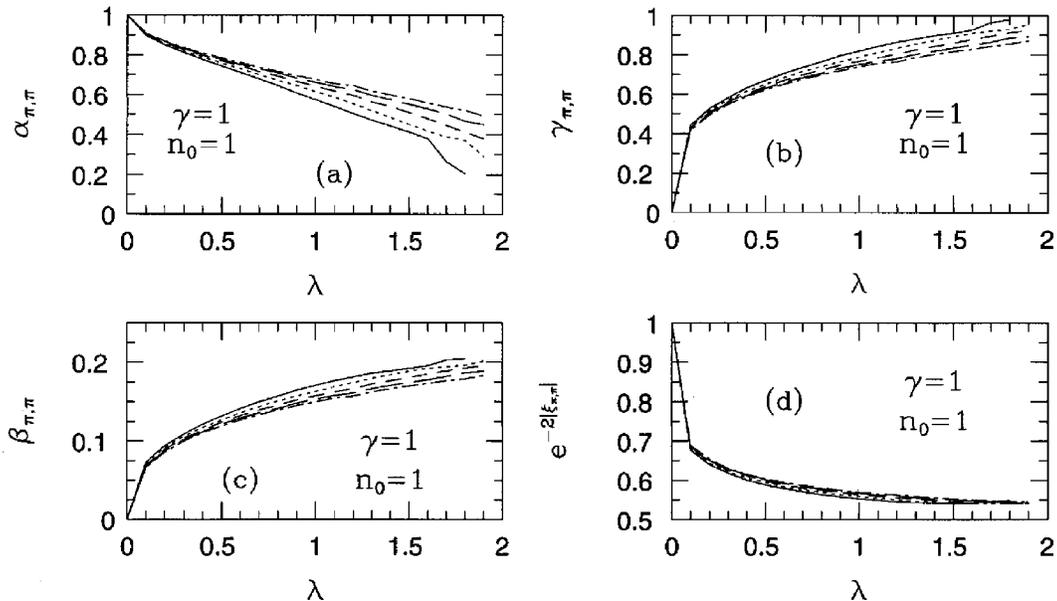

FIG. 2. The parameters of the effective wave function at $\mathbf{k}=(\pi,\pi)$ for $\gamma=1$ and at half-filling $n_0=1$ as a function of the $e$-ph coupling for various Coulomb strengths as $V_c/\omega_0=0,1,2,3,4$ represented by solid, dotted, dashed, long-dashed, and dotted-dashed lines, respectively.

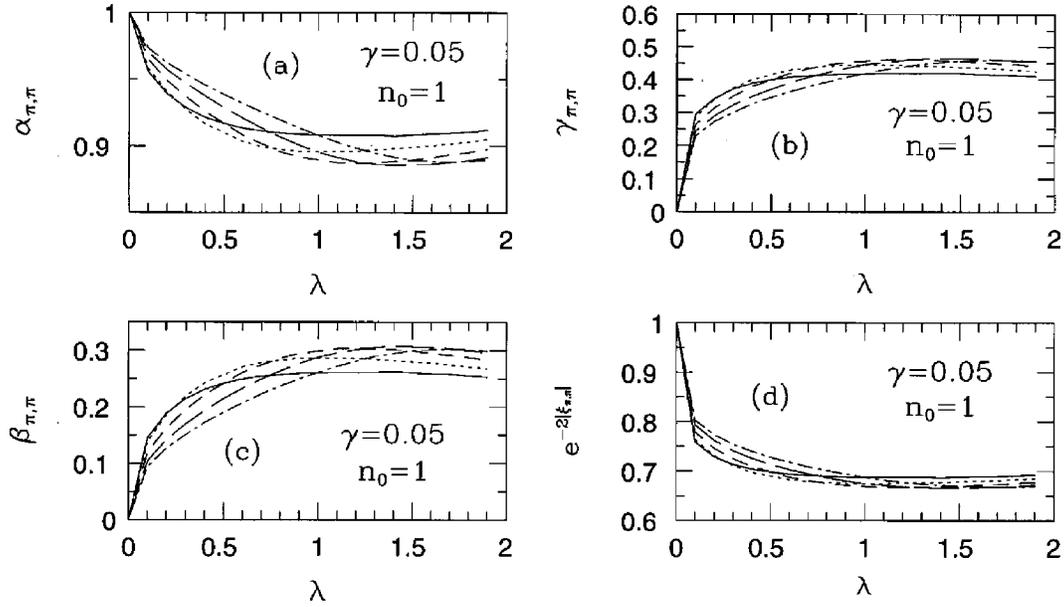

FIG. 3. The same as in Fig. 2 for $\gamma=0.05$ at half-filling $n_0=1$.

As the dependence of this overall picture on the electron concentration is concerned, the first observation we make is that, when $\bar{n}_{\mathbf{k}} \equiv n_0$ is shifted away from half-filling the influence of the Coulomb interaction becomes weaker on all parameters. In addition, the relative strength of the correlated pair excitations (i.e., $\gamma_{\pi,\pi}$) with respect to the two-particle coherent component (i.e., $\alpha_{\pi,\pi}$) becomes weaker as shown in Figs. 4(a,b) for $n_0 \simeq 0.6$. The four-particle correlations as given by $\beta_{\pi,\pi}$ in Fig. 4(c), maintain their negligible strength. We also observe in the same result that the parameters of the two-particle coherent component as indicated in Fig. 4(d) are not too sensitive to changes in the electron concentration in this range.

### 3. Approximate ground-state energy

In Figs. 5(a,b) the ground-state energy difference calculated in reference to the noninteracting limit (i.e., $\lambda=0$) and corresponding to each phonon branch as a function of $\lambda$ is plotted for the same parameter values as the previous figures at half-filling. Note that in this section, we intentionally include the results of both branches in Eqs. (23). To clearly demonstrate the influence of the charge fluctuation correlations, the ground-state energy of the background uniform distribution [i.e., $V_0(\mathbf{k})\bar{n}_\mathbf{k}\bar{n}_{-\mathbf{k}}$] is subtracted in both Figs. 5(a), and 5(b). The first solution obtained for the parameters is identified for each Coulomb strength, by a solid line ($V_c/\omega_0=0$), an open circle ($V_c/\omega_0=1$), an open triangle ($V_c/\omega_0=2$), a solid circle ($V_c/\omega_0=3$), and a solid triangle ($V_c/\omega_0=4$), respectively, in accordance with symbols used in Fig. 1. The second solution is represented by dotted lines, for all Coulomb strengths. At weak $e$-ph coupling strength, a finite positive contribution to the energy is present from Coulombic charge fluctuations. A common feature of all ground-state energy solutions in Fig. 5(a) is that at a fixed Coulomb interaction strength, a slightly lower ground-state energy is obtained with the second branch for coupling strengths $\lambda \lesssim 1$ than with the first branch. In the approximate range $1 \lesssim \lambda$ the first branch yields a lower ground-state energy than the second one. In the transition from one branch to the other no discontinuity is present. In addition to the continuous nature of the transition, a kinklike feature is also present near $\lambda=1$, where the transition is observed. The continuity of the ground-state energy is widely accepted on grounds of direct-diagonalization studies on finite systems[5–8] as well as variational calculations.[9–11] The kinklike feature has also been reported in one-dimensional calculations but it was attributed to the finite-size effects.[8] We also observe, in accordance with Ref. 8 that, as the system parameters are driven into antiadiabatic ranges (i.e., $\gamma \ll 1$) the kinklike feature disappears as shown in Fig. 5(b), and the fluctuations calculated at distinct Coulomb interaction strengths become less viable for the ground-state energy due to the suppression of the dynamical fluctuations.

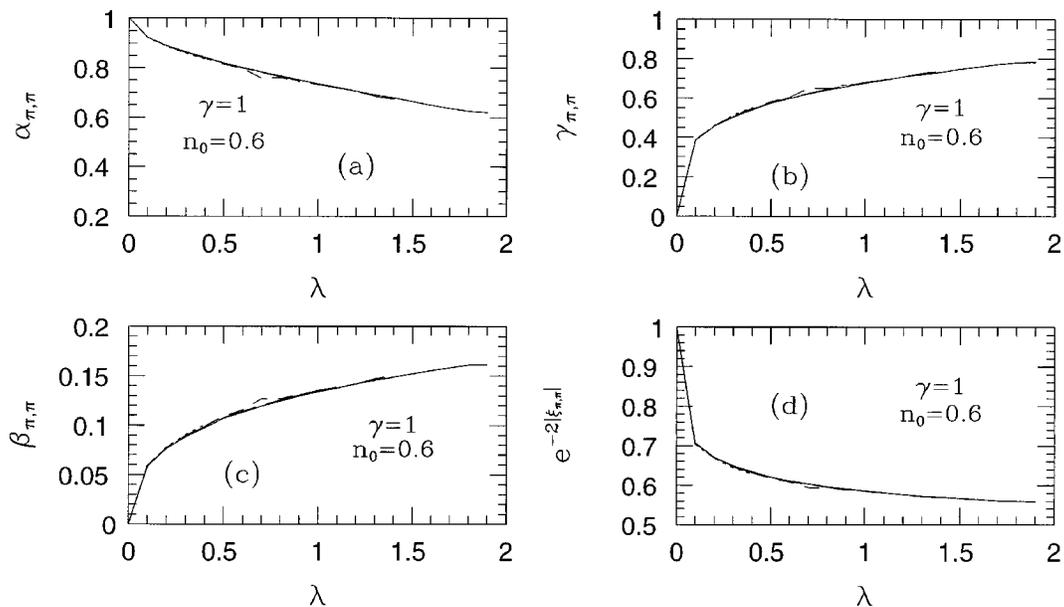

FIG. 4. The same as in Fig. 2 for $\gamma=1$ at electron concentration $n_0=0.6$.

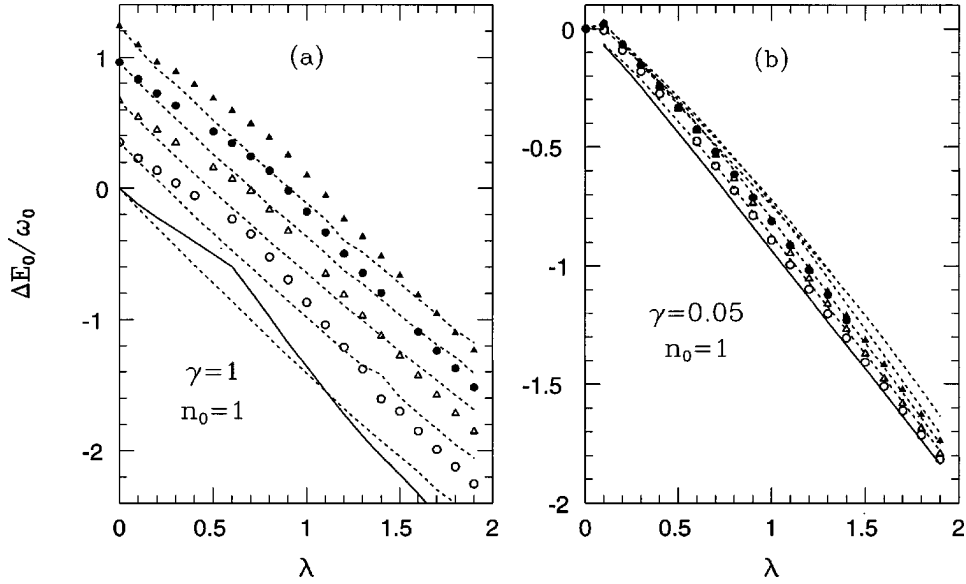

FIG. 5. (a) The ground-state energy difference $\Delta E_0 = E_0(\lambda) - E(\lambda=0)$ as calculated by Eq. (20) for $\gamma=1$ at half-filling and for the two solutions of the wave-function parameters as determined by the values of the phases in Eq. (23). Here, for the second solution the same symbols are used as in Fig. 1 for the same parameter values. Since the first solution and the second one meet on the vertical scale at a value corresponding to a particular value of $V_c$ the second solution for each $V_c$ can be identified easily. For the sake of clarity we thus represent all second solutions with dotted lines. (b) Same as in part (a) for $\gamma=0.05$ at half-filling.

## III. EFFECTIVE CHARGE-TRANSFER AMPLITUDE

It has been shown in the direct-diagonalization calculations on finite systems[5] that the convergence of the intersite charge-transfer amplitude to the conventional Lang-Firsov (LF) limit is weak particularly in the intermediate coupling weakly antiadiabatic regimes. In the conventional LF approach the adibaticity does not play a role in the renormalization of the $t_{\mathrm{eff}}$. The reason behind the independence of $t_{\mathrm{eff}}$ from $\gamma$ is that the standard LF polarons are renormalized only with respect to the lattice site on which the polaron is located; whereas, this approximation is only expected to be manifest in the extreme antiadiabatic strong-coupling limit. On the other hand, the response time scale for the phonon cloud to follow the charge is expected to be a monotonously increasing function of adiabaticity. This implies that in the strongly adiabatic ranges the renormalization of the effective charge-transfer amplitude by the following phonon cloud is expected to be weaker than it is for weakly adiabatic and nonadiabatic ranges. Hence, the localizing effect of the strong $e$-ph coupling should be a function of adiabaticity. This means that $t_{\mathrm{eff}}/t$, as a measure of the kinetic-energy renormalization scale for electrons, is expected to be a monotonously decreasing function when $\gamma$ decreases, which was indeed observed in the numerical calculations of Ref. 5, 7, and 8. In another way of saying it, the expected renormalization of $t_{\mathrm{eff}}$ with respect to $\gamma$ is itself a strong result against the use of the LF approach in the large and intermediate adiabatic ranges and the generality of the argument requires

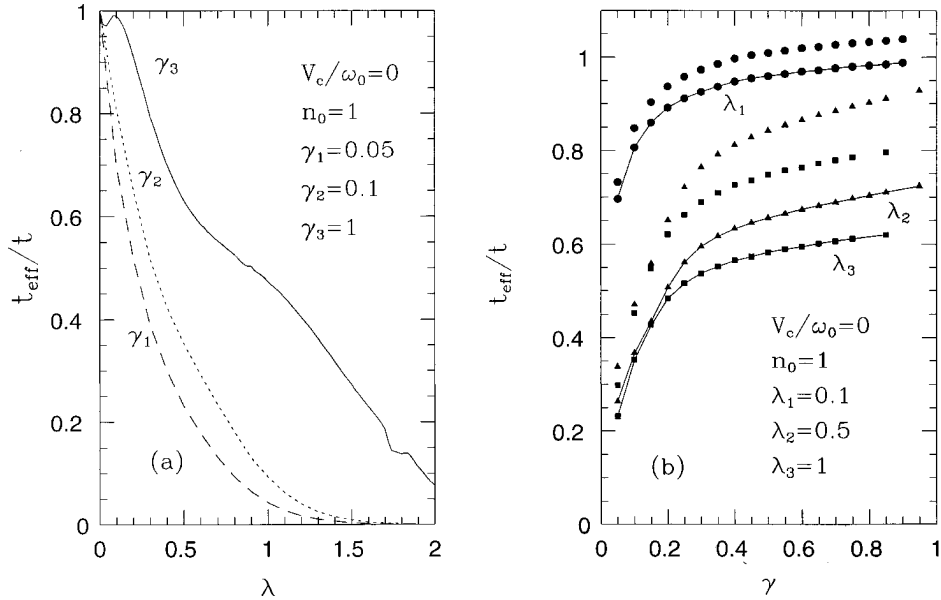

FIG. 6. (a) The effective charge transfer amplitude $t_{\rm eff}/t$ as a function of $\lambda$ for the indicated value of interaction parameters. (b) The adiabaticity dependence of $t_{\rm eff}/t$.

that a similar scenario is expected to hold for the many-body case.

We define the effective charge transfer amplitude $t_{\rm eff}$ using Eq. (21) as

$$t_{\rm eff} = t \langle \psi_\delta^{\rm eff} | \sigma(\mathbf{m},\mathbf{n}) | \psi_\delta^{\rm eff} \rangle. \tag{28}$$

Note that the coherent CDW sector $|\phi_c\rangle$ would have no contribution in Eq. (28) if it was included in the wave function. In Eq. (28), or in its explicit form in Eq. (21), the Lang-Firsov limit would only correspond to $\alpha_\mathbf{k} \equiv 1$, $\gamma_\mathbf{k} = \beta_\mathbf{k} = \xi_\mathbf{k} \equiv 0$, yielding the standard Holstein band reduction $t_{\rm eff} = t\exp(-\lambda/2)$. It can be seen that this limit is unphysical in our dynamical approach here. The reason is that, since all parameters are definite functions of $\lambda$, the limit $\alpha_\mathbf{k} \equiv 1$, $\gamma_\mathbf{k} = \beta_\mathbf{k} = \xi_\mathbf{k} \equiv 0$ would only be obtained if no $e$-ph coupling was present. Hence, deviations from the standard LF approach is an inherent feature of the dynamical approach itself. Since the parameters of $|\psi_\delta^{\rm eff}\rangle$ are known by Eqs. (19) and (22), we can examine Eq. (28) as the $e$-ph coupling constant and the adibaticity are varied. In Fig. 6(a), the coupling constant dependence of the renormalized charge-transfer amplitude is plotted for $\gamma = 0.05$, 0.1, 1. Given the general argument discussed above and the previous results obtained for finite systems, our results in Fig. 6(a) could be qualitatively anticipated, i.e., $t_{\rm eff}$ decreases monotonously with decreasing adiabaticity. To indicate that the adiabaticity dependence is a manifestation of charge fluctuation correlations, Eq. (28) as well as the Lang-Firsov-normalized charge-transfer amplitude $t_{\rm eff}/(te^{-\lambda/2})$ are plotted in Fig. 6(b) as a function of $\gamma$ for $\lambda = 0.1$, 0.5, 1. The connected points with solid circles, solid triangles, and solid squares represent the solution of Eq. (28) for $\lambda = 0.1$, 0.5 and $\lambda = 1$ respectively. The LF-normalized solutions are indicated with the same type of unconnected points for the same $\lambda$ values. The difference between the full and LF-normalized solutions is weaker for small couplings as expected. More importantly, the difference is also a function of the adiabaticity, decreasing monotonously for decreasing $\gamma$. Hence, the qualitative features of Figs. 6(a) and 6(b) reasonably agree with those in Refs. 5, 7, and 8.

## IV. EFFECTIVE ELECTRON-ELECTRON INTERACTION

The effective electron-electron interaction will be calculated from

$$V_{\rm eff}^{e\text{-}e}(\mathbf{k},\omega) = \frac{V_0(\mathbf{k})}{\epsilon(\mathbf{k},\omega)}, \tag{29}$$

where $\epsilon(\mathbf{k},\omega)$ is given by Eq. (25). At half-filling, the calculations are shown for the Coulomb dominated bare interaction in Figs. 7(a,b) for the real and imaginary parts of the inverse dielectric function, Since Re$\{1/\epsilon\}$ is even and Im$\{1/\epsilon\}$ is odd in $\omega$, we only include the positive excitation energies. In the Coulomb dominated region, high-energy excitations across the Fermi surface [i.e., $\omega \sim 2\mu$ and $\mathbf{k} = (\pi,\pi)$] are strongly susceptible to a sharp singularity in the electron density of states where a strong enhancement in the effective $e$-$e$ coupling is observed. In the same limit Im$\{1/\epsilon\}$ has a coherent peak for excitations across the Fermi energy, which is consistent with the known presence of high-energy dynamical CDW fluctuations. In this regime, the quasiparticle screening is inactive and the charge fluctuations are dominated by high-energy processes. We observe that, for weaker bare Coulomb interaction strength the enhancement is also weaker (not shown in Fig. 7). As the bare $e$-ph coupling is increased, the peak position shifts to lower energies due to the quasiparticle band narrowing and the CDW peak amplitude is much less pronounced. In contrast, in the low-energy excitation range (i.e., $\omega \lesssim \mu$), one enters the particle-hole continuum where the screening is active. In this regime, Re$\{1/\epsilon\} < 1$, which suppresses the effective $e$-$e$ coupling below its bare strength.

At the other limit, where the net bare $e$-$e$ coupling is phonon dominated, as shown in [Figs. 8(a,b)], the high en-ergy excitations become incoherent and the coherent CDW instability disappears. Note the presence of a minus sign on the vertical scale in Fig. 8(a) to indicate that the effective $e$-$e$ coupling is attractive ($0<\text{Re}\{1/\epsilon\}$). In this regime, the particle-hole continuum is narrowed from below to intermediate excitation energies where the screening is effective, resulting in a net suppression of the attractive coupling. The limitation of the particle-hole continuum at the low-energy end is dictated by the small polaron formation where a strong enhancement of the attractive coupling is observed. As the bare $e$-ph coupling is increased, the effective polaron mass is strongly enhanced within a low-energy window and the interactions are dominated by low-energy exchange processes. With increasing bare attractive coupling, the low-energy window is compressed to even lower energies, apparently approaching to a $\delta$-like peak at $\omega=0$ for $1\ll\lambda$. For an increasing bare $e$-ph coupling constant, the divergence in the behavior of $\text{Re}\{1/\epsilon\}$ is also consistent with the gradual development of the sharp low-energy peak in $\text{Im}\{1/\epsilon\}$ in Fig. 8(b). We believe that this is an indication of the existence of a very narrow band, itinerant, small (quasilocalized) polarons in this low-energy regime. In the ultimate limit of very large $e$-ph coupling the small polaron band is reduced completely, the effective adiabaticity is strongly decreased and, the effective $e$-$e$ coupling is strongly renormalized signaling a gradual transition from the itinerant, fluctuating low-energy small polaron picture to self-trapped polarons. Since the coupling is strongly attractive, bipolaron bound-state formation is also likely to happen within this range.

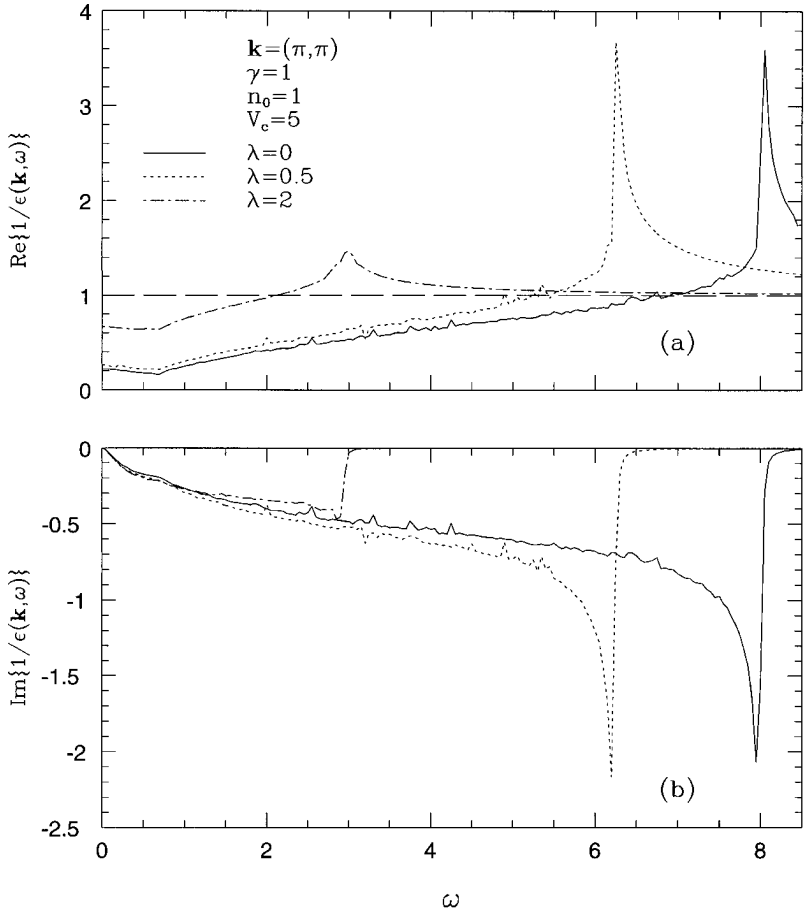

FIG. 7. (a) The real part of the vertex renormalization for the effective $e$-$e$ coupling as a function of the excitation energy $\omega$ in the Coulomb dominated regime at indicated values of the interaction parameters. (b) Same as (a) for the the imaginary part.

Figures 7(a,b) and 8(a,b) confirm the general wisdom[2–4,20] that, the electron self-energy as well as vertex corrections are particularly strong across the Fermi surface both in the high-energy Coulombic and low-energy phonon dominating regimes. To complete the picture at half-filling, the $\mathbf{k}$ dependence of the dielectric function is plotted in Figs. 9(a,b) for $\omega/\omega_0=8.05$, $\lambda=0$, $V_c(\pi,\pi)/\omega_0=4$, and Figs. 10(a,b) for $\omega/\omega_0=0.05$, $\lambda=1.6$, $V_c(\pi,\pi)/\omega_0=0$. These particular $\omega$ values correspond to the vicinity of excitation energies in Figs. 7(a,b) and 8(a,b) where the peak positions are observed. Hence, Figs. 9(a,b) and 10(a,b) give representative samplings of the dielectric function in the extreme high-energy Coulombic and low-energy phonon dominated regimes and where the strongest $\omega,\mathbf{k}$ dependence is expected. In the former [Figs. 9(a,b)] a relatively smooth and dispersionless CDW gap is present on the Fermi surface. Across the Fermi surface at $\mathbf{k}=(\pi,\pi)$ there is an enhancement both in $\text{Re}\{1/\epsilon\}$ and $\text{Im}\{1/\epsilon\}$ indicating the dynamical CDW peak in Figs. 7(a,b). On the other hand, we find in the latter case [Figs. 10(a,b)] that in the presence of a strong attractive coupling the gap fluctuates at very low energies (e.g., $\omega/\omega_0\sim 0.05$), and it is strongly anisotropic on the bare Fermi surface. For instance, at $\mathbf{k}=(0,\pi)$, and at $(\pi,0)$ the $\text{Re}\{1/\epsilon\}$ it is rather flat and narrow with no structure in the imaginary part, whereas across the bare Fermi surface towards $\mathbf{k}=(\pi,\pi)$ it is strongly $\mathbf{k}$ dependent and dynamical with the large dynamical small polaron peak at $\mathbf{k}=(\pi,\pi)$ [see also Figs. 8(a,b)].

An extension of these results to the case away from half filling as well as different values of the bare charge-transfer amplitude also indicate that the $\omega,\mathbf{k}$ dependence of the self-energy and vertex corrections maintain their full validity at a qualitative level. Because of the strong $\omega,\mathbf{k}$ dependence of high-energy excitations in the Coulombic case, the position and the amplitude of the dynamic CDW peak is strongly sensitive to slight changes in the electron concentration. We observed that in the region where low-energy phonon dominated excitations are strong, there is an overall suppression in the magnitude of the low-energy excitations on the Fermi surface as well as at $\mathbf{k}=(\pi,\pi)$ when the concentration is shifted away from the half-filling.

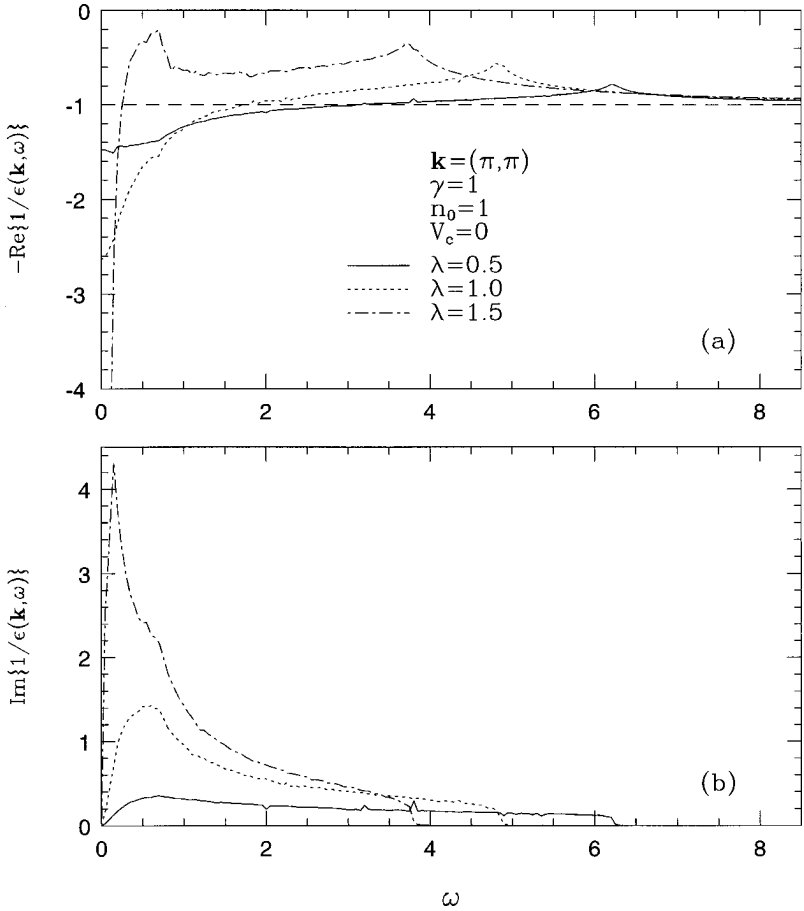

FIG. 8. (a) The real part of the vertex renormalization for the effective $e$-$e$ coupling as a function of the excitation energy $\omega$ for the phonon dominated regime at the indicated interaction parameter values. (b) Same as (a) for the imaginary part.

The density of states on the Fermi surface is strongly dependent on the strength of the charge-transfer amplitude. For $t=0.7$, at half-filling and in the Coulomb dominated case, we observed an order of magnitude enhancement on the Fermi surface in the effective $e$-$e$ interaction. The last example is the extreme phonon dominated region at $t=0.7$ at low energies. There, the previously observed low energy small polaron peak is enhanced and broadened in the vicinity of $\mathbf{k}=(\pi,\pi)$. In addition to that, two dynamical peaks appear in symmetric position at $\mathbf{k}=(0,\pi)$ and $(\pi,0)$. In all examples we examined, relatively more structure is observed in the $\mathbf{k}$ space in the phonon dominated regions than in the Coulombic ones.

The strong sensitivity of the vertex corrections as functions of $\omega,\mathbf{k}$ on the bare interaction parameters and the electron concentration renders the analysis delicate particularly near the instabilities. It has been argued that, in the presence of strong short-range Coulombic or magnetic correlations, the strong enhancement in the phonon-mediated effective attraction can drive the system into superconductivity near the dynamical CDW instability.[20] We believe that this mechanism might be more likely to happen (if it does) in the strongly antiadiabatic ranges in otherwise the same regime where the phonon excitation energies are more compatible with the electronic ones. On the other hand, Coulomb dominated strong coupling antiadiabatic ranges, where the excitations are on the order of bare phonon frequency or smaller with exchange momenta on the order of $\mathbf{k}=(\pi,\pi)$, are also favored by the small polaron formation. Hence the competition in this regime between the superconductivity and quasilocalized polarons, must be decided by the effective adiabaticity as well as the coupling constants. This renders the analysis of the competing effects of the vertex ($\lambda_{\text{eff}}$) and phonon ($\Omega_{\mathbf{k}}$) self-energy against the electron self-energy ($t_{\text{eff}}$) renormalizations to be particularly critical near these instabilities.

## V. RENORMALIZED PHONON SUBSYSTEM

### A. Phonon number distribution

We now examine the distribution of the number of phonons $p(n_{\mathbf{k}})$ in the approximate ground state $|\psi_\delta^{\text{eff}}\rangle$ by

$$p(n_{\mathbf{k}}) = |\langle n_{\mathbf{k}} n_{-\mathbf{k}} | \psi_\delta^{\text{eff}}\rangle|^2. \tag{30}$$

Since $|\psi_\delta^{\text{eff}}\rangle$ is defined in terms of pair excitations we consider $n_{\mathbf{k}} = n_{-\mathbf{k}}$, which allows us to use Yuen's formula,[15,16]

$$\langle n_{\mathbf{k}}, n_{\mathbf{k}}|\mathcal{S}(\{\xi\})|0\rangle = \frac{\sqrt{(2n_{\mathbf{k}})!}}{n_{\mathbf{k}}!} \frac{[\tanh(2|\xi_{\mathbf{k}}|)]^{n_{\mathbf{k}}}}{[\cosh(2|\xi_{\mathbf{k}}|)]^{1/2}}, \quad (31)$$

in the calculation of Eq. (30). We find that

$$\begin{aligned}
\langle n_{\mathbf{k}} n_{-\mathbf{k}}|\psi_\delta^{\text{eff}}\rangle &= \langle n_{\mathbf{k}}, n_{\mathbf{k}}|\mathcal{S}(\{\xi\})|0\rangle \{\alpha_{\mathbf{k}} + \gamma_{\mathbf{k}} \kappa_{\mathbf{k}} \mu_{\mathbf{k}} (2n_{\mathbf{k}}+1) \\
&\quad + \beta_{\mathbf{k}} \kappa_{\mathbf{k}}^2 \mu_{\mathbf{k}}^2 (3n_{\mathbf{k}}^2 + 3n_{\mathbf{k}} + 1)\} \\
&\quad + \langle n_{\mathbf{k}}-1, n_{\mathbf{k}}-1|\mathcal{S}(\{\xi\})|0\rangle \gamma_{\mathbf{k}} \kappa_{\mathbf{k}}^2 n_{\mathbf{k}} \\
&\quad + \langle n_{\mathbf{k}}+1, n_{\mathbf{k}}+1|\mathcal{S}(\{\xi\})|0\rangle \gamma_{\mathbf{k}} \mu_{\mathbf{k}}^2 (1+n_{\mathbf{k}}) \\
&\quad + \langle n_{\mathbf{k}}-2, n_{\mathbf{k}}-2|\mathcal{S}(\{\xi\})|0\rangle \kappa_{\mathbf{k}}^4 n_{\mathbf{k}} (n_{\mathbf{k}}-1) \\
&\quad + \langle n_{\mathbf{k}}+2, n_{\mathbf{k}}+2|\mathcal{S}(\{\xi\})|0\rangle \mu_{\mathbf{k}}^4 (n_{\mathbf{k}}+1) \\
&\quad \times (n_{\mathbf{k}}+2). \quad (32)
\end{aligned}$$

Using Eq. (32) and (31), the phonon number distribution in Eq. (30) is plotted for different values of $\lambda$ and $\gamma$, $n_0$ and $V_c/\omega_0$ at $\mathbf{k}=(\pi,\pi)$ in Figs. 11(a–d). The values of the coupling constants are chosen sufficiently below and sufficiently above the critical crossover of the two solutions near $\lambda \simeq 1$ in Fig. 5(a) so that $p(n_{\mathbf{k}})$ is calculated using the first solution for $\lambda_1$ and $\lambda_2$ and the second one for $\lambda_3$. A common feature of Figs. 11(a–d) is that, for sufficiently small (i.e., $\lambda = \lambda_1$), the phonon probability distribution is always the largest at $n_{\mathbf{k}}=0$. As $\lambda$ increases, the maximum value is smoothly shifted towards finite number of phonons and the overlap with the vacuum state decreases. As the system is driven into antiadiabatic ranges, as shown in Fig. 11(b), there is an overall decrease in the dynamical charge fluctuation correlations where the phonon distribution is narrower and the overlap with the vacuum is strongly increased. A comparison between Figs. 11(a) and 11(b) indicates that there is a delicate competition between $\gamma$ and $\lambda$ to determine the shape of the probability distribution. The decreasing $\gamma$ tends to compress the distribution towards $n_{\mathbf{k}}=0$ by increasing the vacuum component. On the other hand, a weak (i.e., $\lambda=\lambda_1,\lambda_2$) but increasing $\lambda$ broadens the distribution and attempts to shift it away from the vacuum, where it fights against the stabilizing effect of the decreasing $\gamma$. Whereas, if $\lambda$ is strong (i.e., $\lambda=\lambda_3$), the increasing $\lambda$ cooperates with the decreasing $\gamma$ to stabilize the coherent polaron formation as indicated by the increasing $n_{\mathbf{k}}=0$ component in $p(n_{\mathbf{k}})$. We identify the cooperation of increasing $\lambda$ and decreasing $\gamma$ as the correct route to the Lang-Firsov limit in which the dynamical component of the probability distribution very strongly overlaps with the vacuum where the phonon statistics is driven by the dominating coherent part.

A similar competition is observed in Figs. 11(a) and 11(c) between the $e$-ph and the Coulomb interactions, as well as in Figs. 11(a) and 11(d) for different electron concentrations. When $\lambda$ is weak, increasing $\lambda$ competes with the stabilizing effects of Coulomb interaction or reduced electron concentration. When $\lambda$ is strong, it cooperates with them to stabilize the coherent polaron formation. We observe that the overall picture here is also consistent with the results of de Mello and Ranninger in Ref. 5.

It should be noted that the nonclassical structure of $p(n_{\mathbf{k}})$ is entirely a manifestation of the dynamical fluctuations. The fluctuating part given by $|\psi_\delta^{\text{eff}}\rangle$ in Eq. (15) of the true ground-state wave function does not support any structural changes [i.e., $\langle \psi_\delta^{\text{eff}}|Q_{\mathbf{k}}|\psi_\delta^{\text{eff}}\rangle = \langle \psi_\delta^{\text{eff}}|P_{\mathbf{k}}|\psi_\delta^{\text{eff}}\rangle = 0$ as also enforced by Eqs. (18)]. Hence, the decomposition of the wave function in the correlation space also enables one to examine the dynamical and static parts of the distribution function independently. The true probability distribution is obtained by a convolution between the dynamical and static coherent sectors of the wave function. The static coherent sector yields the nonfluctuating Poisson distribution, which is not addressed in this paper.

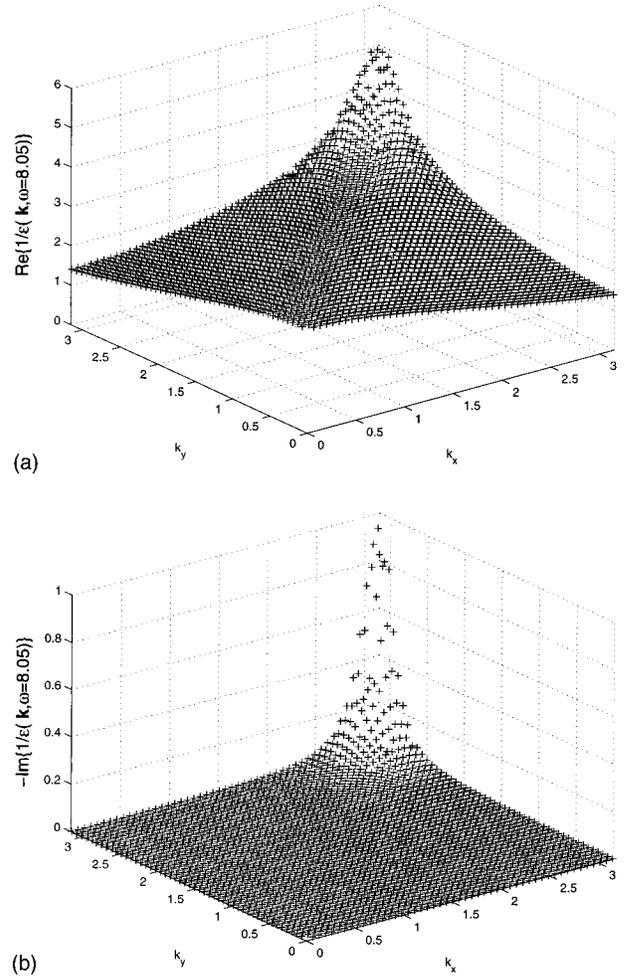

FIG. 9. (a) The real part of the vertex renormalization for the effective $e$-$e$ coupling in the Coulomb dominated regime at the peak value $\omega/\omega_0 = = 2\mu/\omega_0 = 8$ in $\mathbf{k}$ space for $V_c/\omega_0 = 4$, $\lambda = 0$, $\gamma = 1$ and $n_0 = 1$. (b) Same as (a) for the imaginary part (note the negative sign on the vertical scale).

### B. Renormalized frequency of vibrations

In principle, the phonon frequency renormalization should be calculated by finding the corresponding effective phonon Hamiltonian for which the dynamical polaron wave function in Eq. (15) is the *lowest eigenstate*. This would be a tedious, but relatively straightforward inverse eigenproblem if we could write the operator in Eq. (15) in the form of an *invertible* unitary operator acting on the phonon vacuum state. In the following, we will present our results instead, using the RPA where the phonon self-energy $\Pi(\mathbf{k},\omega)$ is calculated by

$$\Pi(\mathbf{k},\omega) = V_0(\mathbf{k})\mathcal{P}(\mathbf{k},\omega), \qquad (33)$$

where $\mathcal{P}(\mathbf{k},\omega)$ is the electron polarization given by Eq. (26). We will present our results for the phonon dominated regime without Coulomb interaction. Hence $V_0(\mathbf{k})=-\lambda$. The RPA is known to yield compatible results to the self-consistent ME calculations[2,4] in the relatively weak-coupling constant ranges $\lambda \lesssim 0.5$ whereas it strongly overestimates the dynamical phonon softening for $0.5<\lambda$ as compared to more reliable QMC simulations.[2] In the conventional RPA the renormalized phonon frequency is given by

$$\Omega_\mathbf{k} = \sqrt{\omega_\mathbf{k}^2 + 2\omega_\mathbf{k}\Pi(\mathbf{k},\omega_\mathbf{k})}, \qquad (34)$$

where bare electron Green's functions are normally used in the calculation of $\mathcal{P}(\mathbf{k},\omega)$. Using Eq. (34), we plot in Fig. 12 the renormalized phonon frequency $\Omega_\mathbf{k}$ in the RPA (thin solid lines) as a function of $\lambda$ for $\gamma=0.3, 0.4, 1$ and for no Coulomb repulsion. The ME calculations (dotted lines) and QMC results (with error bars) of Ref. 2 for $\gamma=1$ are also included for comparison. It is known that the conventional RPA overestimates the charge fluctuation correlations due to neglected corrections of the self-consistent renormalizations in the electron self-energy and the coupling constant.[2–4,20] This is reflected in an unbounded negative increase of the phonon self-energy, which in turn derives the renormalized phonon frequency into an instability for the intermediate and strong-coupling ranges $1 \lesssim \lambda$.

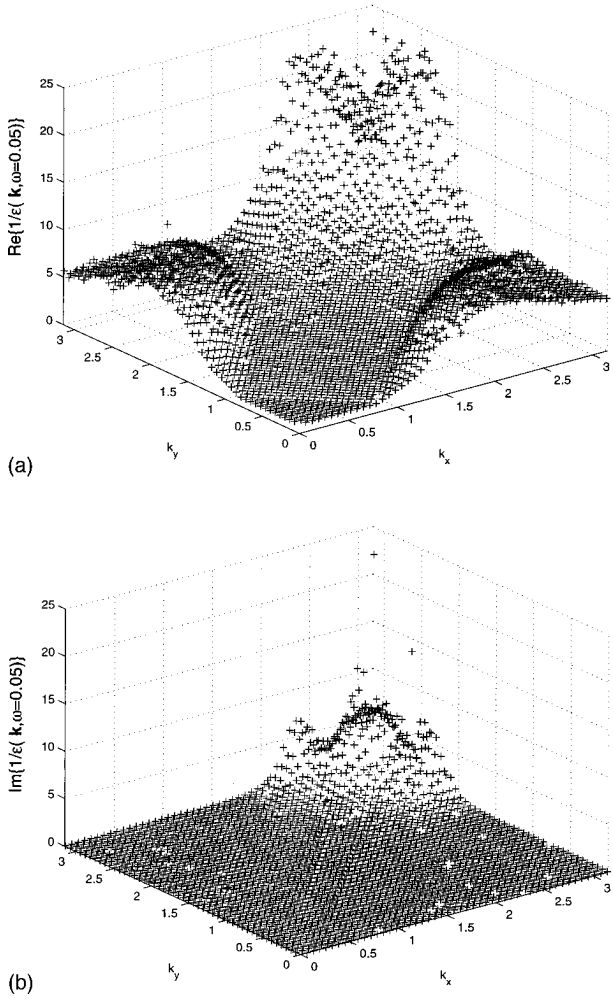

FIG. 10. (a) The real part of the vertex renormalization for the effective $e$-$e$ coupling in the phonon dominated regime at the peak value $\omega/\omega_0=0.05$ in $\mathbf{k}$ space for $V_c/\omega_0=0$, $\lambda=1.6$, $\gamma=1$ and $n_0=1$. (b) same as (a) for the imaginary part.

If the vertex corrections are properly included, in the attractive case, the effective $e$-ph coupling constant $\lambda_\text{eff} = \lambda\,\text{Re}\{1/\epsilon(\mathbf{k},\omega)\}$ is suppressed for high frequency excitations due to the charge screening effect and is enhanced in the low frequency range due to the small polaron formation [see the coupling constant renormalization in Sec. IV Fig. 8(a,b)]. On the other hand the electron self energy is also reflected upon the renormalization of the charge transfer amplitude $t_\text{eff}$ of which the band narrowing effect, according to Fig. 6(a), is to derive the system into an effectively nonadiabatic range. Hence a physically more relevant calculation should properly include *both* corrections which is suggested by replacing $\Pi(\mathbf{k},\omega_\mathbf{k}) \to \Pi_\text{eff}(\mathbf{k},\Omega_\mathbf{k})$ in Eq. (34) where the latter is calculated with $t\langle\sigma\rangle_\text{LF} \to t_\text{eff}$ where $t_\text{eff}$ is now given by Fig. 6(a), and, with $\lambda \to \lambda_\text{eff}$ where $\lambda_\text{eff} = \lambda\,\text{Re}\{1/\epsilon(\mathbf{k},\omega_\mathbf{k})\}$ is calculated in Fig. 8(a). The self-consistent solution of

$$\Omega_\mathbf{k} = \sqrt{\omega_\mathbf{k}^2 + 2\omega_\mathbf{k}\Pi_\text{eff}(\mathbf{k},\Omega_\mathbf{k})}, \qquad (35)$$

which we term as the corrected RPA (CRPA), is technically different from those calculations using finite lattice and electron degrees of freedom where it is numerically feasible to maintain the self-consistency from the beginning.[4] The solution of the CRPA is depicted in Fig. 12 with the thick solid line as a function of the bare coupling constant $\lambda$. In the solution of CRPA, we were not able to beyond $\lambda \simeq 1.6$ due to an unstability in the numerical calculations in Eq. (35). Nevertheless, the agreement with the QMC results for a reasonably large range of $e$-ph coupling clearly indicates the importance of the vertex as well as the self-energy corrections in the antiadiabatic strong-coupling case. The picture can be made more transparent if one divides the $\lambda$ range in Fig. 12 by imaginary lines into the weak-coupling $\lambda \lesssim 0.5$, intermediate-coupling $0.5 \lesssim \lambda \lesssim 1.2$, and strong-coupling $1.2 \lesssim \lambda$ sectors and compare the $\gamma=1$ RPA solution where such renormalizations are not present with the $\gamma=1$ CRPA solution where they are included. In the weak sector, the phonon softening is weak and typical excitation energies are on the order of bare phonon frequency where the charge screening effects weakly suppress the coupling constant (i.e., $\text{Re}\{1/\epsilon\}<1$). By the weak screening in this sector, further softening of phonons is slightly delayed to the larger coupling strengths. In the intermediate range, the charge fluctuations become important where the electron self-energy and vertex corrections compete to determine the phonon softening. This can be qualitatively understood by the following argument. As $\lambda$ is increased in the intermediate range, the band narrowing effect of the electron self-energy corrections tend to oppose further softening, but in the intermediate sector the phonon frequency is already sufficiently softened and the low-energy excitations slowly start dominating as a precursor of the fluctuating polaronic regime where the large low-energy vertex corrections enhance the effective coupling constant $1<\text{Re}\{1/\epsilon\}$. Hence, more softening is observed. On the other hand, in the third sector at relatively large coupling con-stants, the outcome of the competition between the electron self-energy and vertex corrections is decided by the bare adiabaticity parameter $\gamma$. At this point, it is necessary to go back and examine the renormalization of the charge-transfer amplitude in Fig. 6(a) for various values of $\gamma$. For intermediate and large values of $\lambda$, the band reduction is opposed by the suppression factor $\exp(-4\xi_\mathbf{k})$ in Eq. (21) arising from the strong presence of the two-particle coherent (i.e., $0 < \xi_{\pi,\pi}$, $\alpha_{\pi,\pi} < 1$) and, the two-particle pair excitations (i.e., $0 < \gamma_{\pi,\pi}$) in the ground state. The net effect of the coherent two-particle pair excitations is to slow down the rapid reduction of the electron band as $\lambda$ increases. The influence of this factor has also been noticed in the variational calculations in the intermediate and strong couplings as well as intermediate and low excitation energies in the phonon spectrum.[21,22] We observe in Fig. 6(a) that, this effect is visible for $\gamma = 1$ by the presence of a bulge near $\lambda = 0.6$ and the decrease of $t_{\text{eff}}/t$ for increasing $\lambda$ is much slower for the larger values of $\gamma$. This implies that, a smaller $\gamma$ yields a more rapid band reduction, resulting in a stronger suppression of the charge fluctuations. In the strongly antiadiabatic regime, the increasing $e$-ph coupling cooperates with the strong nonadiabaticity [as also observed in Fig. 11(b)] and the phonon softening is completely destroyed. This is indicated in Fig. 12 by the thin solid lines corresponding to $\gamma = 0.4$ and 0.3. On the other hand, for larger $\gamma$, the phonon softening can continue in the presence of marginal charge fluctuations. For instance, for $\gamma = 1$ and for the CRPA solution, as $\lambda$ is increased further, the charge fluctuations decrease, leading into a finite saturation regime where the phonon softening is relatively unchanged with $\lambda$.

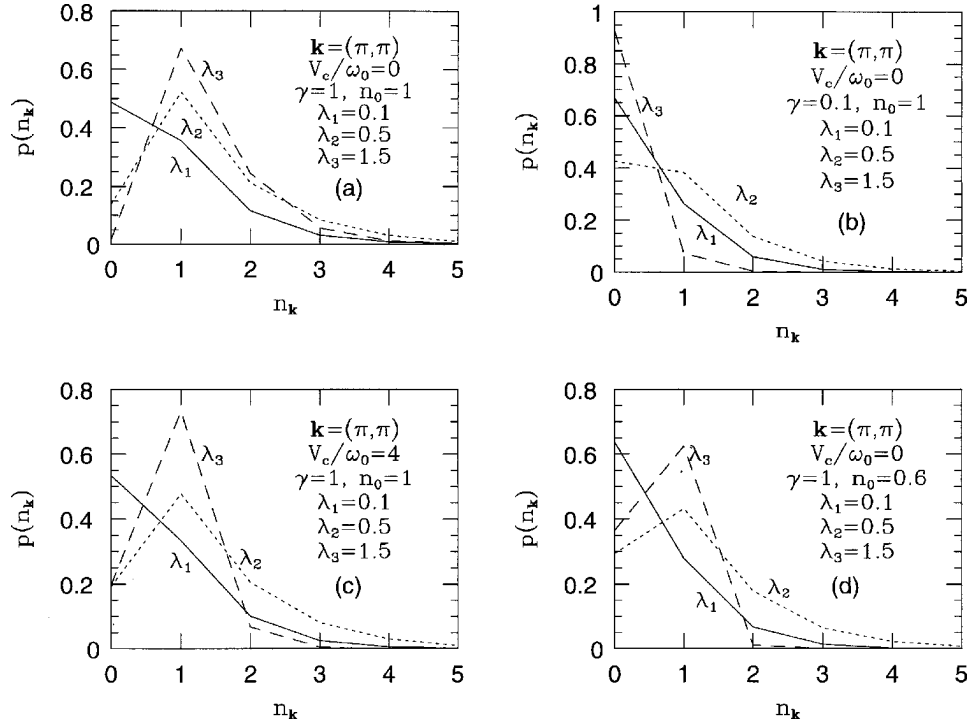

FIG. 11. The dynamical phonon distribution in the effective wave function for the indicated values of the parameters.

## VI. CONCLUSIONS

In this work, we improved and extended the dynamical charge fluctuation based effective wave-function scheme of our previous work in Ref. 13 to the normal state in the two-dimensional Holstein-Hubbard model in the intermediate interaction ranges. In particular, the possibility of representing the effective wave function in the decoupled subspaces of $n$-phonon cumulant correlations is exploited and applied to the first two cumulants of the polaron wavefunction. The differences of this approach from the diagrammatic phonon correlator technique of Ref. 12 as well as the standard Lang-Firsov approaches are emphasized by showing that the numerically observed weak convergence to the LF theory in the strong-coupling antiadiabatic limit is inherently built in this model. With the effective cumulant approximation, one is able to construct an effective many-body wave function and compare the results at a qualitative level with the recent numerical studies on direct diagonalization, QMC, and variational approaches. The effective wave function provides a clear picture of the dynamical coupling of the correlated phonon pair fluctuations to those in the CDW. In this respect, we consider the current work as a possible dynamical many-body extension of these studies.

As far as the general polaron problem is concerned, the decoupled nature of the effective wave function in the cumulant correlation space might be a promising tool to understand the properties of the polaron ground state at a deeper level. This procedure also decouples the static coherent sector from the dynamical fluctuating part of the wave function. In this article we took this as an advantage to study the dynamical sector independently. The authors believe that the possible improvements of this extended LF-like approach can be done in two directions. At first one can realize that, the true ground state [as suggested by the multiphonon scattering operator $\sigma(\mathbf{m},\mathbf{n})$] has corrections to the coherent part even at the dynamical level, and, the true ground-state wave function includes a dynamically shifted mixture of coherent states of the phonon coordinate $Q_\mathbf{k}$ and momentum $P_\mathbf{k}$. The static contribution is a $Q_\mathbf{k}$ coherent state, which is precisely what we called $|\phi_c\rangle$ in this article. In $|\phi_c\rangle$ we have neglected these pure dynamical corrections, although a more rigorous treatment should also embody those effects self-consistently. A second means of improvement is in the understanding of $|\psi_\delta\rangle$ itself. At this point, some formal difficulties arising from the formulation of the unitary generators of the $m$-phonon cumulant correlations for $3 \leq m$ have to be overcome. The cumulant correlation corrections for $3 \leq m$ also depend heavily on the corrections to the Landau-Fermi liquid picture. The reason behind this is that higher cumulants in $|\psi_\delta\rangle$ are more susceptible to deviations from the standard assumption of Gaussian density fluctuation correlations in the Landau-Fermi liquid. This assumption was indeed used in the calculation of Eqs. (11). In this respect, these two corrections to phonon as well as fermion statistics should be attacked simultaneously in a more refined self-consistent frame. Possible advances made in this direction might reveal the importance of these deviations and might also shed light on the likely presence of the not-completely-understood strongly nonlinear self-trapping regime both in the Coulombic high-energy and phonon dominated low-energy sectors.

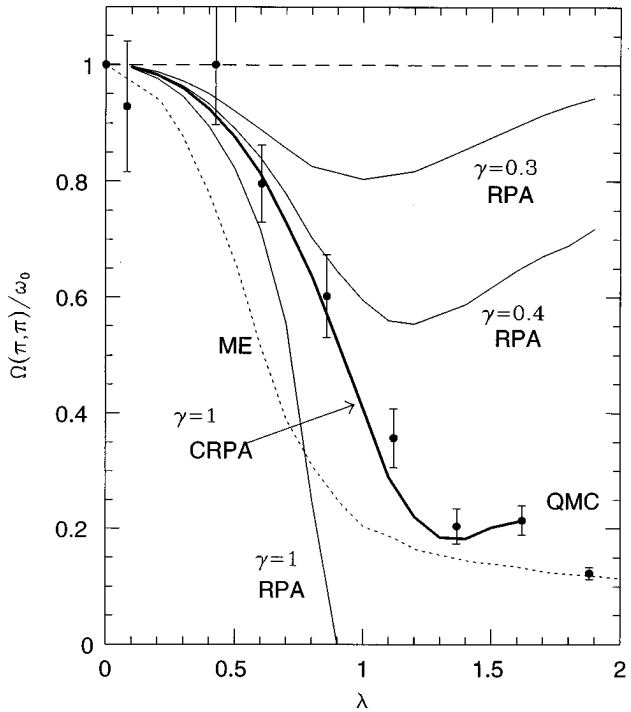

FIG. 12. The phonon softening comparatively studied for the (corrected) random-phase approximation, Migdal-Eliashberg as well as quantum Monte Carlo calculations of Ref. 2 for the indicated parameter values at $\mathbf{k}=(\pi,\pi)$ and at half-filling.

Although the revival of the Holstein-Hubbard model in the past 15–20 years was heavily stimulated by the progress in high-temperature superconductivity, we did not enter into such discussions in this article. Using an oversimplified model, it was suggested in Ref. 13 that the low-temperature $T_c$-dependent phonon anomalies observed in certain Cu-O–based compounds might be connected with the dynamical vibrational fluctuations self-consistently coupling to the polaronic charge fluctuations in the superconducting phase. It should be noted that a more realistic model for high-temperature superconductors is suggested by the Holstein–$t$-$J$ model in the presence of strong Coulomb correlations with the electron concentration being slightly shifted away from half-filling where the vibrational fluctuations strongly couple in a self-consistent frame to charge but also spin fluctuations in the Cu-O planes. One then has to incorporate all self-energy and vertex corrections in the Coulomb dominated regime, both for the fluctuations in the charge and spin degrees. Hence, one possible direction to take in the superconducting phase is to examine the Holstein–$t$-$J$ model within the (charge and spin) fluctuation-based effective cumulant approach presented here.

### ACKNOWLEDGMENTS

T.H. is grateful M. Arai, C. H. Booth and, in particular, to N. Bulut for helpful and stimulating discussions. Both authors are indebted to V. A. Ivanov for the exchange of crucial ideas. M.Y.Z. is grateful to TÜBİTAK (Scientific and Technical Research Council of Turkey) and Bilkent University for support and hospitality.